\documentclass[preprint,12pt]{elsarticle}
\usepackage[caption=false]{subfig}
\usepackage{amsmath}
\usepackage{mathtools}
\usepackage{url}
\usepackage{pdflscape}
\usepackage[]{placeins}
\usepackage{multirow}
\usepackage[]{threeparttable}

\def\rot{\rotatebox[origin=c]}

\journal{NUCL INSTRUM METH A}
\begin{document}
\begin{frontmatter}
\title{The radiation field in the Gamma Irradiation Facility GIF++ at CERN}

\author[a,b]{Dorothea~Pfeiffer\corref{cor1}}
\author[c,a]{Georgi~Gorine}
\author[d]{Hans~Reithler}
\author[a,e]{Bartolomej~Biskup}
\author[a]{Alasdair~Day}
\author[a]{Adrian~Fabich}
\author[a]{Joffrey~Germa}
\author[a]{Roberto~Guida}
\author[a]{Martin~Jaekel}
\author[a]{Federico~Ravotti}

\cortext[cor1]{Corresponding author: dorothea.pfeiffer@cern.ch}
 
\address[a]{CERN, CH-1211 Geneva 23, Switzerland}
\address[b]{European Spallation Source (ESS ERIC), P.O. Box 176, SE-22100 Lund, Sweden}
\address[c]{Ecole Polytechnique F\'{e}d\'{e}rale de Lausanne(EPFL), CH-1015 Lausanne, Switzerland}
\address[d]{	RWTH Aachen University, 52062 Aachen, Germany}
\address[e]{Institute of Experimental and Applied Physics, Czech Technical University in Prague, 16636 Prague 6, Czech Republic}

\begin{abstract}
The high-luminosity LHC (HL-LHC) upgrade is setting now a new challenge for particle detector technologies. The increase in luminosity will produce a particle background in the gas-based muon detectors that is ten times higher than under conditions at the LHC. The detailed knowledge of the detector performance in the presence of such a high background is crucial for an optimized design and efficient operation after the HL-LHC upgrade. A precise understanding of possible aging effects of detector materials and gases is of extreme importance. To cope with these challenging requirements, a new Gamma Irradiation Facility (GIF++) was designed and built at the CERN SPS North Area as successor of the Gamma Irradiation Facility (GIF) during the Long Shutdown 1 (LS1) period. It features an intense source of 662 keV photons with adjustable intensity, to simulate continuous background over large areas, and, combined with a high energy muon beam, to measure detector performance in the presence of the background. The new GIF++ facility has been operational since spring 2015. In addition to describing the facility and its infrastructure, the goal of this work is to provide an extensive characterization of the GIF++ photon field with different configurations of the absorption filters in both the upstream and downstream irradiation areas. Moreover, the measured results are benchmarked with Geant4 simulations to enhance the knowledge of the radiation field. The absorbed dose in air in the facility may reach up to 2.2 Gy/h directly in front of the irradiator. Of special interest is the low-energy photon component that develops due to the multiple scattering of photons within the irradiator and from the concrete walls of the bunker. 
\end{abstract}


\end{frontmatter}

\section{Introduction}
\label{sec:intro}

The Gamma Irradiation Facility (GIF)~\cite{Agosteo} at CERN~\cite{CERN}, the European Organization for Nuclear Research, was extensively used from 1997 until its closure in 2014 for the characterization of particle detectors. Located in the former CERN SPS West Area, the facility played in particular an important role in testing large area muon detector systems and components for the Large Hadron Collider (LHC)~\cite{LHC}. In this facility, detectors could simultaneously be exposed to the photons from a $^{137}$Cs source and to a high-energy muon beam. Although from 2005 onwards only the caesium source was available, the GIF continued to be fully exploited all year round by a wide community of users. The high-luminosity LHC (HL-LHC) upgrade~\cite{HL-LHC} is setting now a new challenge for particle detector technologies. The increase in luminosity will produce a particle background in the gas-based muon detectors that is an order of magnitude higher than under present conditions at the LHC, hence detector tests at accordingly higher rates are required. The detailed knowledge of the detector performance in the presence of such a high background is crucial for an optimized design and efficient operation after the HL-LHC upgrade. A precise understanding of possible aging effects of detector materials and gases is also of extreme importance. To cope with these challenging requirements, a new Gamma Irradiation Facility (GIF++) was designed and built during the Long Shutdown 1 (LS1) period at the CERN SPS North Area as successor of the Gamma Irradiation Facility (GIF)~\cite{Jaekel}. Whereas CERN was responsible for the construction of the facility and the procurement of the irradiator, a comprehensive user infrastructure was provided within the framework of the FP7 AIDA project~\cite{Aielli}. The new GIF++ facility offers two separated irradiation areas and has been operational since spring 2015.

The goal of this note is to provide an extensive characterization of the GIF++ photon field with different configurations of the absorption filters in both the upstream and downstream irradiation areas. Measured dose rates are benchmarked with Geant4~\cite{Geant4a} simulations to enhance the knowledge of the radiation field. Of special interest is the low-energy photon component that develops due to the multiple scattering of photons within the irradiator and from the concrete walls of the bunker.  A detailed description of the shielding of the GIF++ bunker and its optimization with the help of Monte Carlo simulations is in preparation~\cite{FLUKA}.

\section{The GIF++ facility}
\label{sec:facility}

\subsection{Layout of the facility}
\label{subsec:Layout of the facility}
\begin{figure}[htb]
  \centering
  \includegraphics[width=0.85\textwidth]{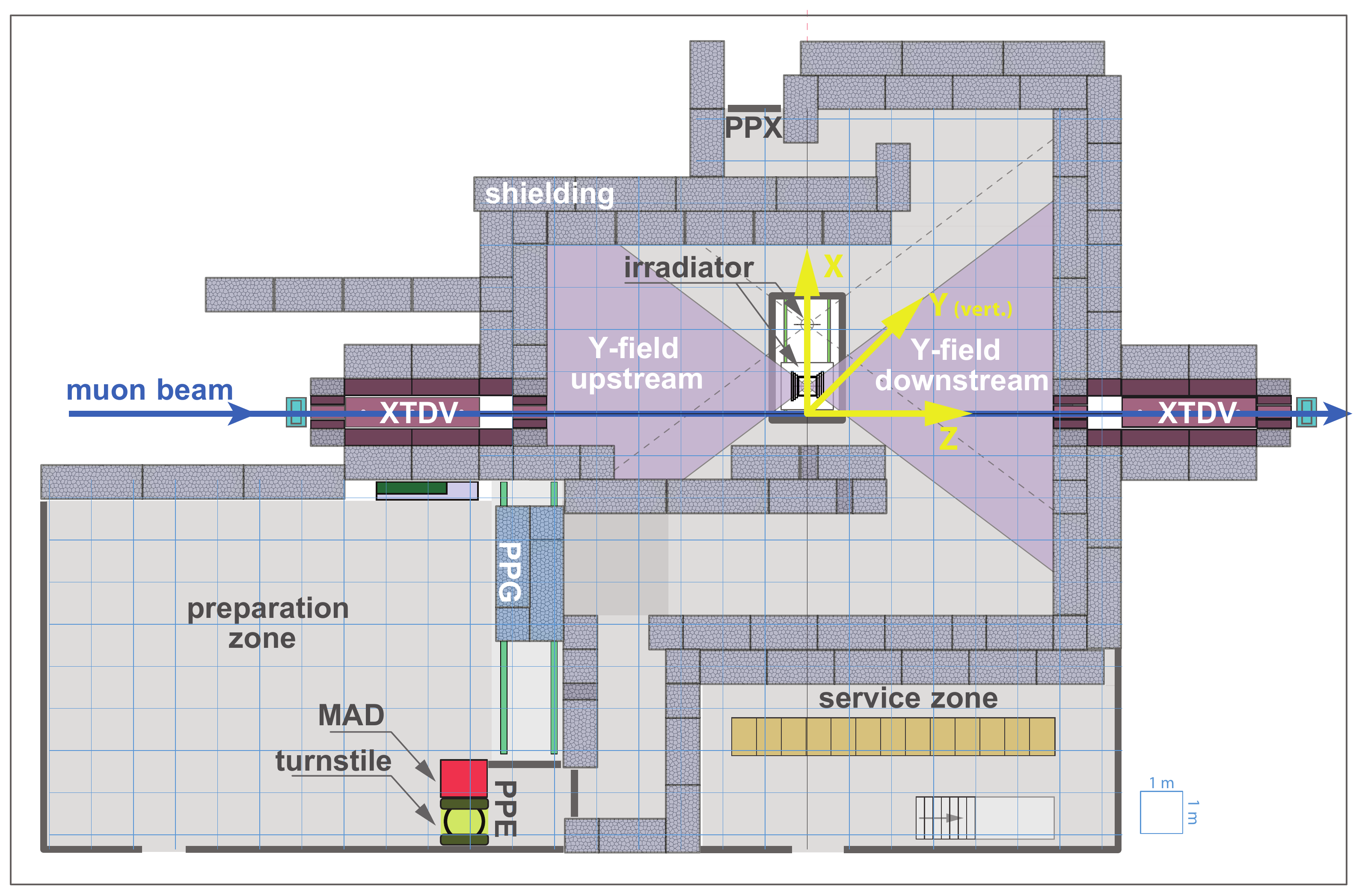}%
  \caption{Floor plan of the GIF++ facility with entrance doors MAD (material access door), PPG (personal protection gate), PPE (personal protection entrance), PPX (personal protection exit). When the facility downstream of the GIF++ takes electron beam, a beam pipe is installed along the beam line (z-axis) between the vertical mobile beam dump (XTDV). The irradiator can be displaced laterally (its center moves from x~=~0.65~m to 2.15~m), to increase the distance to the beam pipe.}
   \label{fig: facility}
\end{figure}

Focused on the characterization and understanding of the long-term behavior of large gas-based particle detectors, GIF++ combines a $^{137}$Cs source\footnote{Source activity: 14.9 TBq in November 2011, 13.9 TBq in March 2015, 13.5 TBq in March 2016.} with two sets of adjustable filters to vary the intensity, and a high-energy muon beam (100 GeV/c) from the secondary SPS beam line H4 in EHN1. The $^{137}$Cs isotope was chosen instead of $^{60}$Co due to its long half-life of 30.08 years, leading to a smaller decrease of the photon rate over the expected lifetime of this facility. Further, the typical energy of the neutron-induced background radiation at LHC experiments like CMS~\cite{Mueller} approximately matches the energy spectrum of the $^{137}$Cs source, composed of the primary 662 keV photons and lower energetic scattered photons. The layout of the GIF++ facility and the used coordinate system are shown in Figure~\ref{fig: facility}. 

\subsection{High-energy reference beams}
\label{subsec:High-energy reference beams}
Located near the end of the beam line H4, GIF++ is the main user of this beam line for six to eight weeks per year. In addition, the facility receives parasitic muon beam halo for 30-50$\%$ of the SPS operation time. The muon beam is generated as a secondary beam from the primary SPS proton beam on a production target. The spectrometer of the H4 beam line allows the beam line user to choose the nominal momentum of the secondary beam with a maximum momentum of 400 GeV/c. The muon beam in the energy range of 57-100$\%$ of the nominal beam energy is mainly generated by the decay of pions and kaons that are produced in the primary target. Depending on the beam line settings, either the full bandwidth or an energy spread reduced to the percent level is transported to the GIF++ area. The spill structure of the muon beam follows the primary proton beam structure, which has a spill of 4.8 seconds with a close to flat distribution. Depending on the SPS cycling for other users (e.g. LHC), one spill arrives about every 30 seconds on average. With the full acceptance of the beam line, the intensity can be up to 10$^4$ muons per spill, which is limited by radiation protection aspects. The secondary beam can also be adjusted for hadrons and/or electrons. For about five weeks per year an electron beam is granted to main users whose experiments are located downstream of the GIF++ area. During this time, an evacuated pipe is installed all along the beam line in the GIF++ bunker, but the photon source in GIF++ can still be used. In order to increase the distance to the beam pipe when needed, provision is made to shift the whole irradiator laterally. The lateral distribution of the muon beam at the GIF facility depends on several factors like the final focusing, which can be varied within a large range. In case of parasitic use, the settings chosen by the primary beam line user upstream determine the intensity and energy. As a general rule, the core of the muon beam covers a surface of 10~cm x 10~cm, containing half of the muon beam. The remaining part of the muon beam, the beam halo, is spread over a footprint of about 1~m$^2$.

\subsection{Irradiator and filter system}
\label{subsec:Irradiator and filter system}
\begin{figure}[htbp]
\centering
\captionsetup[subfloat]{justification=centering}
\subfloat[Irradiator with angular correction filters\label{fig: irradiator}]{
\includegraphics[width=.2386\textwidth]{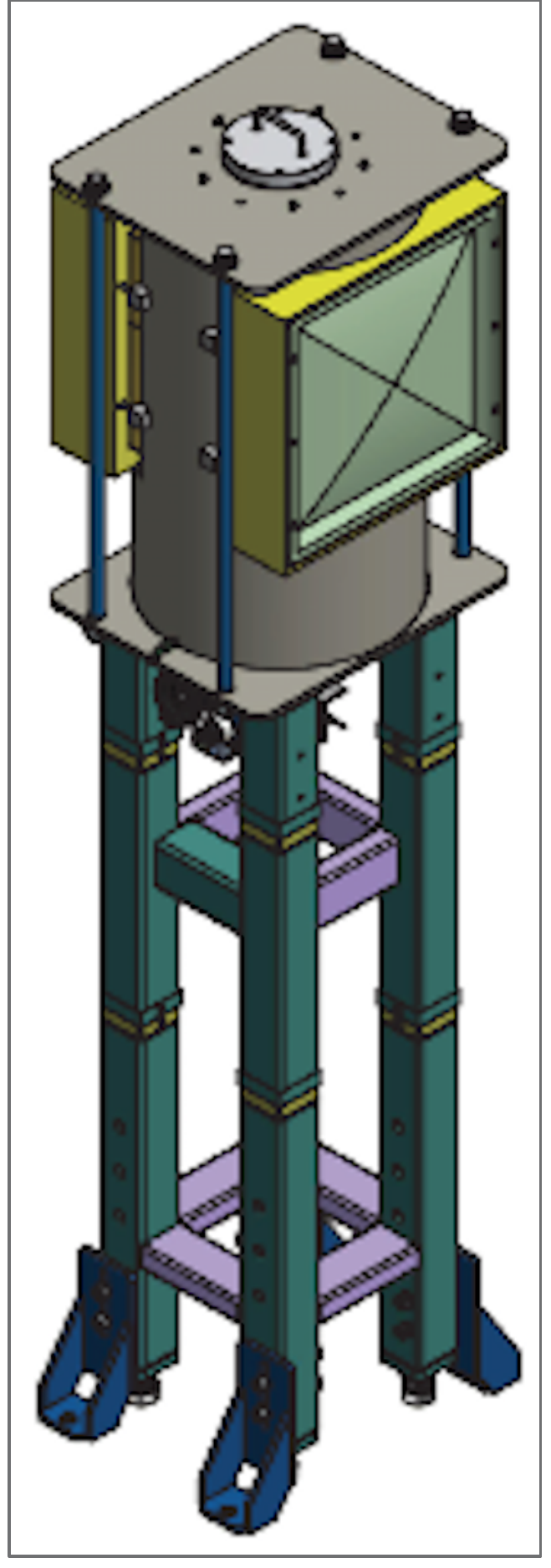}%
}
\subfloat[Irradiator with filter system\label{fig: filters}]{
\includegraphics[width=.40\textwidth]{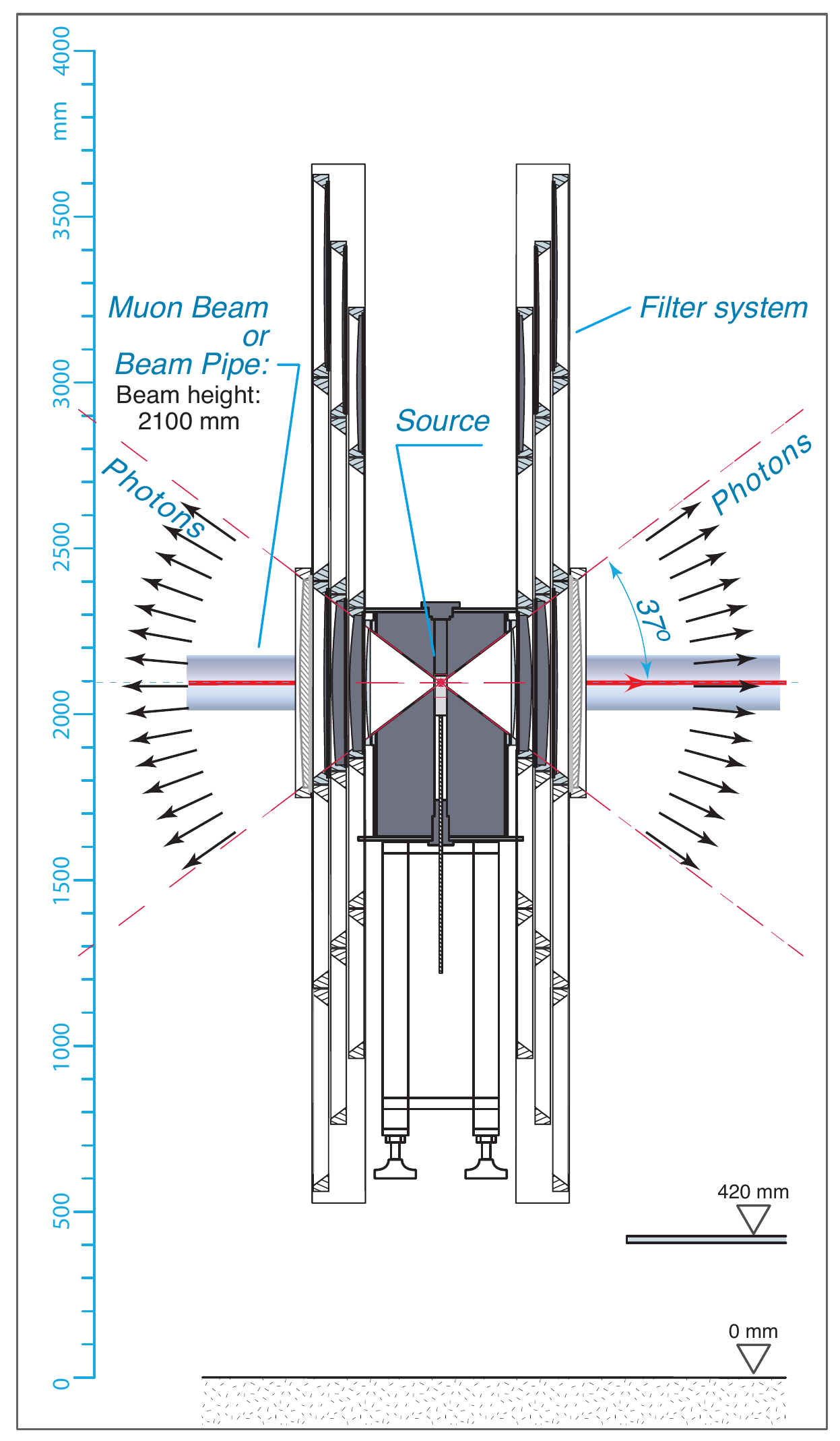}%
}
\subfloat[Nominal attenuation factors\label{fig: filter_factors}]{
\includegraphics[width=.2835\textwidth]{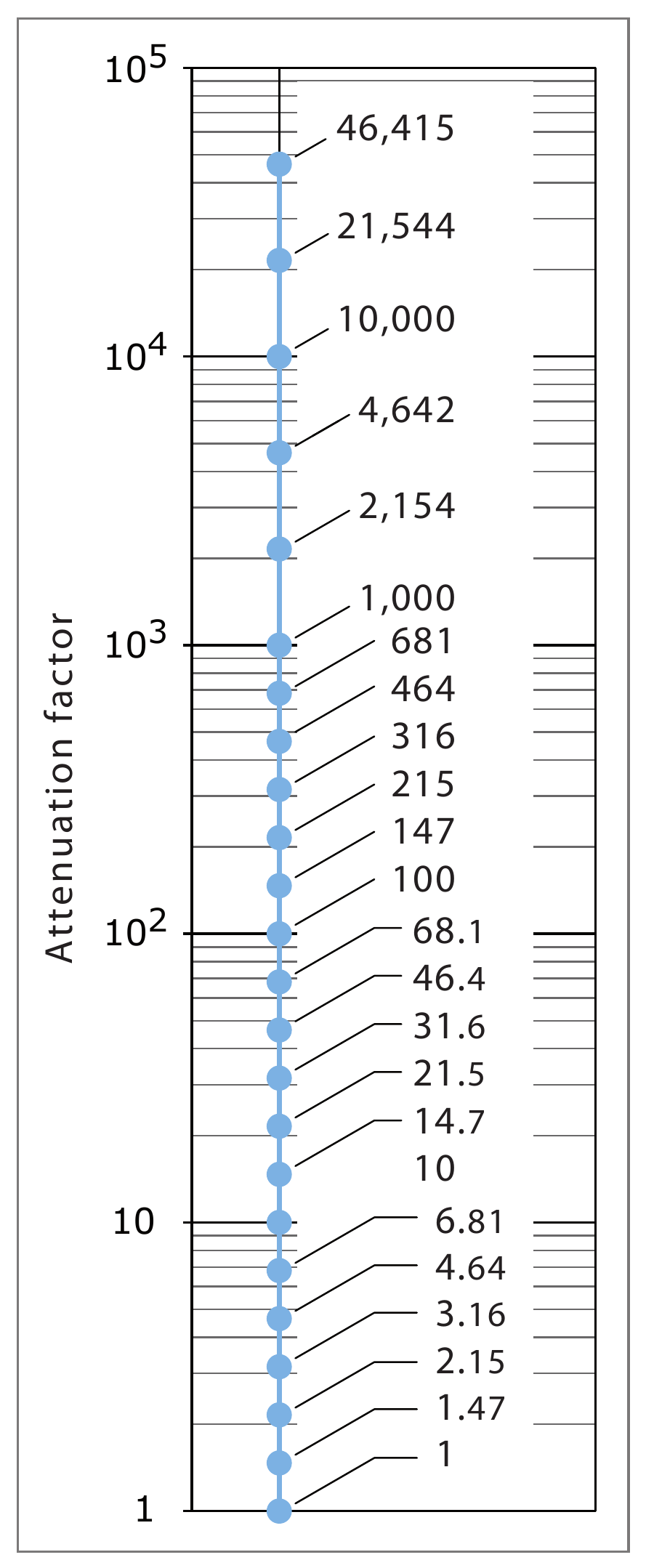}%
}
\caption{(a) Schematic drawing of the GIF++ irradiator with angular correction filters and independent filter systems at both sides. (b) On both sides a set of independently movable and remotely controlled attenuation filters and collimator frames permits to vary the intensity of emitted photons. (c) Each set of 3x3 movable attenuation filters, with 27 combinations of filters, leads to 24 different attenuation factors between 1 and 46415. The two collimators have vertically and horizontally an opening of $\pm$~37$^\circ$ with respect to the beam axis.}
\label{fig: Filters and filter factors}
\end{figure}

Gamma irradiation is however available throughout the whole year, except during short maintenance periods. The 100~m$^{2}$ GIF++ bunker is about 5~m high and has two independent irradiation zones (named as upstream and downstream in Figure~\ref{fig: facility}), making it possible to test simultaneously several real size detectors, with a size of up to several square meters, as well as a broad range of smaller prototype detectors and electronic components. The GIF++ irradiator and its filter systems are depicted in Figure~\ref{fig: Filters and filter factors}. The irradiator has been developed in cooperation with the Czech company VF a.s.~\cite{VF}. The caesium source can be moved from the garage position at the bottom of the support tube inside the shielded receptacle to the irradiation position at the top of the tube.  

With two $\pm$~37$^\circ$ panoramic collimators, the irradiation zone covers a large part of the bunker area, both in the downstream and upstream regions. As shown in Figure~\ref{fig: irradiator}, both outlets of the irradiator are equipped with a lens shaped angular correction filter to provide a uniform photon distribution over a plane, as needed for flat large area detectors. Embedded inside a common enclosure, two complete and independent attenuation systems are available (Figure~\ref{fig: filters}), each consisting of an array of 3x3 convex lead attenuation filters, to fine tune the photon flux for each irradiation field individually. A collaboration with the RWTH Aachen~\cite{RWTH} took care of design and production of the angular correction and attenuation filters. Mounted on aluminum support plates, the filters are positioned inside steel frames, as collimators, and connected to counterweights moving on the side of the irradiator. Three planes of filters (A, B, C), with three filters per plane are installed. The filters have the nominal attenuation factors 1 (A1,B1,C1), 1.5 (B2), 2.2 (C2), 4.6 (C3), 10 (A2) and 100 (A3, B3). The total attenuation factor, from the three filter layers, can be set both via a dedicated control panel, as well as via the GIF++ control system. In total, 24 different nominal attenuation factors between 1 and 46415 can be selected according to the 27 possible combinations. The factors were chosen to be nearly equidistant, on a logarithmic scale, over the first three orders of magnitude, as illustrated in Figure~\ref{fig: filter_factors}. The irradiator and the two filter systems are mounted on rails, so that the whole assembly can be moved in the direction transverse to the muon beam.

\subsection{Angular correction filter and attenuation filters: Adjustment of current}
\label{subsec:modification of current}
Without the angular correction filter, the GIF++ source would be approximately a point source. The angular correction filter, made of steel, is shaped in such a way, that the $\frac{1}{r^2}$ dependence of the photon current is replaced by a uniform current in each xy plane. The $\pm$~37$^\circ$ opening of the irradiator collimators allows horizontally and vertically angles $\theta$ (angle between the incoming photon and the surface normal of the filter) from 0$^\circ$ in the center of the filter to 46.8$^{\circ}$ in the corners of the collimator. To reach a uniform current, the photons have to be attenuated with a factor of cos$^{-3}$($\theta$~-~46.8$^\circ$). That means that the photons in the center of the filter are attenuated by a factor of cos$^{-3}$(46.8$^\circ$), whereas the photons in the corners of the collimators are not attenuated. Each angle of incidence corresponds to a different path length traversed in the filter. The actual attenuation of the photon depends on this path length and the linear attenuation coefficient $\mu$ of the material\footnote{After traversing a length of $x$~cm in a material, the intensity $I$ of a beam of mono-energetic photons with original intensity $I_0$ amounts to I=I$_{0}$e$^{-\mu x}$. The following linear attenuation factors for photons with E=662~keV were used for the design of the filters: 0.116 mm$^{-1}$ for Pb, 0.020 mm$^{-1}$ for Al, 0.053 mm$^{-1}$ for Fe.}. The thickness of the angular correction filter varies for each angle of incidence in such a way, that the desired attenuation is reached for all 662~keV photons. Here attenuation can either mean that the photon has lost energy, or that it has been fully absorbed. For photons with lower energies, created due to scattering in the source capsule and the collimator of the irradiator, the angular correction filter leads to a less uniform current. 

\begin{figure}[htbp]
 \centering
 \includegraphics[width=.785\textwidth]{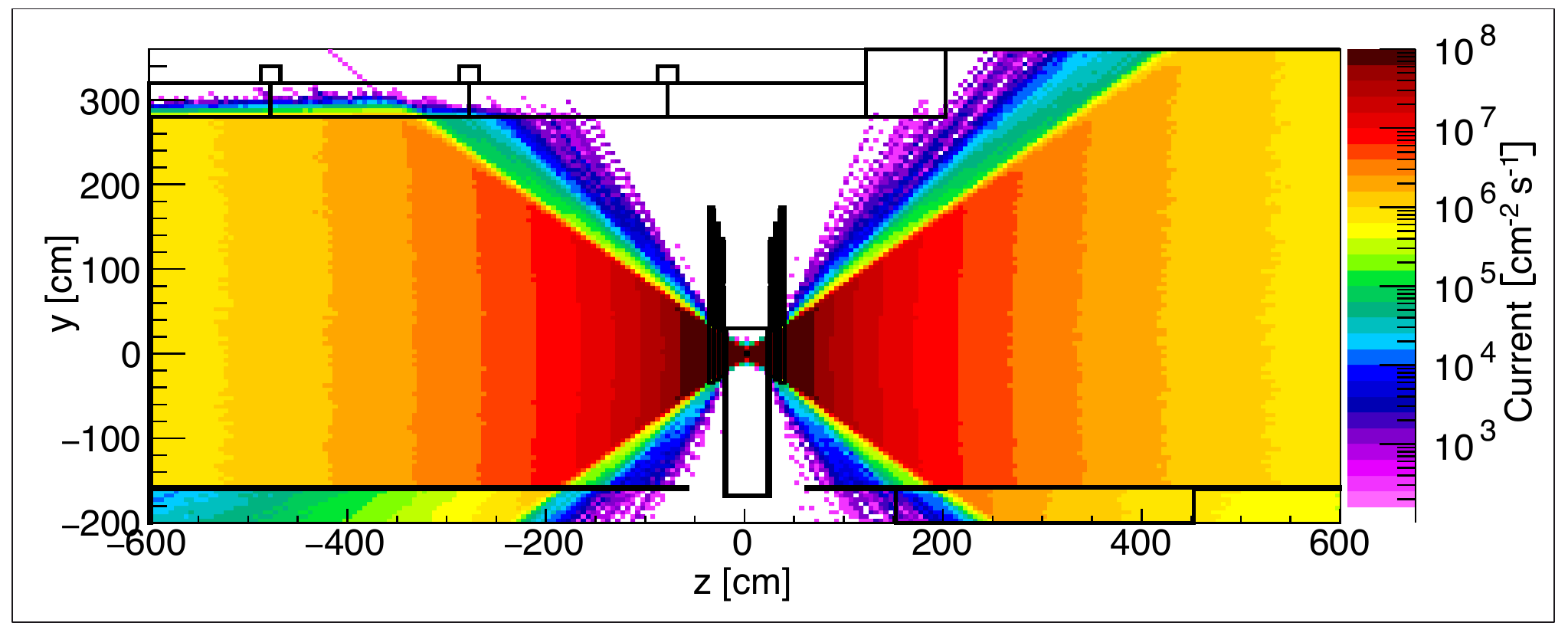}%
 \caption{Photon current in the vertical plane through the source (yz plane) at x~=~0.65~m; attenuation filters at factor 1. With angular correction filters, the current of 662~keV photons is made uniform in xy planes. }
 \label{fig: current_662keV_y}
\end{figure}

For the attenuation filters, on the other hand, the attenuation factor does not depend on the angle of incidence of the photons. The convex face of the attenuation filters is shaped in such a way that at every point of the filter, photons from the source traverse the same thickness of material and hence undergo the same attenuation. The nominal attenuation factor is again defined as attenuation of 662~keV photons; photons with lower energy are attenuated to a larger degree. Due to the presence of lower energy photons, the effective total attenuation over the whole spectral range is lower than the nominal attenuation. The exact value will depend on the spectral sensitivity of the detector used, as well as on the presence of other objects in the bunker. Figure~\ref{fig: current_662keV_y} shows the simulated current of 662~keV photons in the GIF++ facility in the yz plane for an attenuation factor 1, i.e. unattenuated. This is thus the highest current obtainable at GIF++. With the help of the angular correction filters, the current depends only on the distance from the irradiator along the z-coordinate and is uniform in all xy planes. The corresponding xz view of the current of 662~keV photons can be found in Figure~\ref{fig: Current_DS_US_x_600_667}.

\subsection{Area monitoring}
\label{subsec:monitoring}
Ambient dose equivalent and ambient dose equivalent rates in the GIF++ zone are monitored by the Radiation Protection Group (RP) of CERN's Health and Safety unit using the RAdiation Monitoring System for the Environment and Safety\footnote{RAMSES comprises more than 400 radiation monitors (not including environmental monitors) and, in conjunction with two other systems, provides radiation monitoring throughout CERN's accelerators, experimental and service areas.} (RAMSES)~\cite{RAMSES}.  In the GIF++ facility, two RAMSES Induced Activity Monitors\footnote{PTW 32006 air filled 3~L ionisation chambers from ptw.de with a calibrated measurement range of 5 $\mu$Sv/h - 500 mSv/h} are installed at fixed locations. One is attached to the bunker wall upstream and the other downstream at a similar height. The monitors are connected to signal conditioning electronics contained within a rack mounted monitoring station. The monitoring stations are designed in such a way that they perform their function independently of their supervisory system. They are equipped with an uninterruptible power supply that allows them to assure their function without external electrical power for at least two hours. In the case of GIF++, RAMSES generates hardwired interlocks to the access control system to block access to the zone whilst the source is exposed. RAMSES allows remote on-line supervision of all the measured variables as well as data logging and long-term archiving for off-line data analysis and reporting. The RAMSES installation is thus part of the safety infrastructure of the zone, but in addition provides radiation monitoring information for the users of the facility.

Within the framework of the EU-funded AIDA project, the Institute For Nuclear Research And Nuclear Energy in Sofia (Bulgaria), provided a Radiation Monitoring (RADMON) readout system for the monitoring of the integrated dose delivered by the photon source. The RADMON sensor was developed at CERN~\cite{RADMON} as a flexible and convenient solution for radiation monitoring, and is currently being used by several experiments at CERN and external facilities. Various dosimeters with different sensitivities and dynamic ranges can be mounted on the RADMON Integrated Sensor Carrier(ISC) to monitor radiation fields with wide ranges of ionizing dose and particle fluence. Starting in October 2015, the RADMON readout system has been fully implemented within the GIF++ bunker. In total 12 RADMON sensors can be freely positioned in the downstream and upstream area of the facility. The sensors are read out automatically by the GIF++ Control System and data are made available via software through the DIP communication protocol. In contrast to the RAMSES monitors, the RADMONs are not a safety system but a dedicated radiation monitoring system for the GIF++ users.

\section{Setup of simulations and measurements}
\label{sec:measurement and simulations}

\subsection{Units}
\label{subsec:units}
In the present paper, the air kerma\footnote{The acronym kerma stands for \underbar{k}inetic \underbar{e}nergy \underbar{r}eleased in per unit \underbar{ma}ss. The kerma is defined as $K=\frac{dE_{tr}}{dm}$, where dE$_{tr}$ is the sum of the initial kinetic energies of all the charged ionizing particles liberated by uncharged ionizing particles in a material of mass dm. The absorbed  dose is defined as $D=\frac{d\bar{\epsilon}}{dm}$, with $d\bar{\epsilon}$ the mean energy imparted to matter of mass $dm$. Absorbed dose and kerma are numerically identical, provided that electron equilibrium is attained and radiative losses are negligible~\cite{ICRP119}. This is basically the case in the air of the radiation zone of GIF++.} $K_a$ is used to quantify the absorbed dose in air. Kerma and absorbed dose have both the unit Gy and are closely related, but the kerma is usually more convenient to calculate~\cite{IAEA}. The ambient dose equivalent\footnote{The ambient dose equivalent H$^*$(10), an operational quantity for area monitoring, is measured in Sv and is used for strongly penetrating radiation like photons above 15~keV.} H$^*$(10), the unit used by area radiation monitors, can be converted to air kerma $K_a$ with the help of conversion factors~\cite{ICRP74}.

The particle fluence is defined as $\Phi=\frac{dl}{dV}$, the sum of all the particle trajectories $dl$ per unit volume $dV$~\cite{ICRP119}. The unit of the fluence is m$^{-2}$. The fluence rate or flux is defined as $\phi=\frac{d\Phi}{dt}$ and has the unit of m$^{-2}$s$^{-1}$. The deposited energy of a particle in matter is proportional to the particle flux, therefore flux is the adequate unit when dealing with detectors for calorimetry. 

On the other hand, current is a measure of the net number of particles crossing a \textit{flat surface} with a well-defined orientation. The unit of current is m$^{-2}$s$^{-1}$ and thus identical to the unit of flux. Current is meaningful in cases where particles are counted without any interest in their interactions. For example, a fully efficient infinitesimal thin detector would record each photon entering the detector independent of the actual energy deposited by the particle. 

For the GIF++ it was decided, that the typical detector tested in the facility comes rather close to this model, and thus current and not flux was used as design quantity for the angular correction filter. In a directed radiation field flux and current are identical for normal incidence to the surface. At all other angles the flux is higher by a factor of cos$^{-1}$($\theta$), with $\theta$ as the angle between incoming particle and surface normal of the scoring surface~\cite{Huhtinen}.

\subsection{Simulation setup}
\label{subsec:simulation setup}
Fluence and current in the GIF++ photon field were simulated with Geant4 10.0 and the G4EmLivermorePhysics physics list. For this purpose, the whole simulated facility was partitioned using a mesh of 5~cm x 5~cm x 5~cm scoring cubes. In each of the cubes, the photon fluence and photon current binned in 100~keV bins were calculated\footnote{The primitive scorers \textit{G4PSCellFlux} and \textit{G4PSFlatSurfaceCurrent} were used~\cite{PS}. A production cut for gammas of 28.5 keV in lead and 2.14 keV in concrete was used in the simulations.}. Additionally, for the measurement locations listed in Table~\ref{table: dose}, the spectral distribution of the photons was simulated in 1~keV bins. From the simulated fluence, the air kerma $K_a$ was calculated using the conversion factors published in reference~\cite{Veinot}. For the albedo studies, the interactions of photons of different energies and angles of incidence with the materials present in the facility (lead, steel, concrete, aluminum) were simulated.  

\subsection{Measurement locations}
\label{subsec:measurement locations}
Dose measurements were carried out in the GIF++ facility in March 2015, December 2015 and March 2016 in the positions shown in Table~\ref{table: dose}. D stands for positions downstream, U for positions upstream and I for a position outside the irradiation area. The origin of the GIF++ coordinate system (Figure~\ref{fig: facility}) is defined in x and y by the position of the beam line and in z, along the beam line, by the center of the source.

\subsection{RAMSES and Automess gamma probe 6150AD-15 measurements}\label{subsec:automess and ramses measurements}
The RAMSES data shown here have been taken in March 2015, prior to the installation of any user equipment. Additionally a gamma probe\footnote{The gamma probe 6150AD-15~\cite{gamma probe}, from Automess, contains a Geiger-M\"uller counting tube as detector and is used in combination with the dose rate meter 6150AD~\cite{probe manual}.} was used to measure the ambient dose H$^*$(10). The useful dose rate range of the probe is 1 mSv/h - 9.99 Sv/h. The manufacturer calibrated the probe with a 333~kBq $^{137}$Cs calibration source. Within the useful dose rate range, a linearity deviation of the measured intensity of up to $\pm$~10$\%$ compared to the calibration is permitted. The energy range extends from 65~keV to 3 MeV. The nominal angular range is $\pm$~45$^\circ$ around the preferential direction perpendicular to the axis of the tube. The combined energy and directional dependence shows a maximum deviation of up to $\pm$~40$\%$ for all energies and directions compared to 662~keV photons from $^{137}$Cs arriving in the preferential direction~\cite{gamma probe, probe manual}. The ambient dose equivalent H$^*$(10) was converted to air kerma $K_a$ using the factors described in reference~\cite{Units}.  

\subsection{RADMON measurements}\label{subsec:radmon measurements}
In December 2015 RADMON~\cite{RADMON} measurements were conducted in the downstream area, and in March 2016 in the upstream area. All the RADMON have been equipped with an on-board temperature sensor and two RADFETs for ionizing dose measurement\footnote{As RADFET a REM 250 from radfet.com and a LAAS 1600 from laas.fr were used.}. The choice was driven by the need for a device with sensitivity in the range of several mGy, but capable to integrate doses up to the tens of kGy range. As a passive integrating sensor, the RADMON gives as information the total cumulative dose. Normally it serves as device for monitoring radiation levels, but can also be used as a dose rate dosimeter. Due to the very small active area of the sensor, the length of the cables and the precision of the electronics readout, the minimum detectable dose rate at GIF++ was found to be in the order of 10 mGy/h. Consequently, a consistent dose rate measurement at GIF++ can be achieved only by operating the sensors at an adequately high dose rate, or by allowing a sufficient integration time to obtain a measurable dose increase. In order to fully test the RADMON performance in the GIF++ photon field and to check the correct functioning of the readout system, several calibration measurements have been performed. The presence of other experimental equipment in the zone lead to slightly modified RADMON positions compared to the survey in March 2015, and also affected the measurements.

The experimental data have been collected in two runs. The measurements in the downstream area were performed over two weeks of irradiation in December 2015. In that period the attenuation factors were varied between 1 and 100. The source duty cycle and the different attenuation factors were taken into account when calculating the dose rates. Measurements in the upstream area were conducted in March 2016 over three days of operation, with the source always on and the attenuation factor always equal to 1. The measured dose rate was automatically corrected for temperature variations, although the variation during this test stayed within $\pm$~5$^\circ$C, which is lower than the minimum needed to increase the dose readout by one bit. Additionally, the losses related to annealing were negligible, due to the relatively short exposure time. Since April 2016, the GIF++ bunker has been equipped with an air conditioning system that allows the control of the temperature and thus limits the temperature variations. The annealing on the other hand has to be taken into account in long-term measurements or after periods of long inactivity of the sensor.

\section{Results}\label{sec:results}

\subsection{Measurements and simulations of the filter attenuation factors}\label{subsec:attenuation}
An attenuation of 1 means that photons emerging from the angular correction filter are not further attenuated, and is subsequently called \textit{fully open}. The maximum attenuation factor of 46415 is called \textit{fully closed}. Table~\ref{table: attenuation measured} shows that for the lower attenuation factors, the dose attenuation measured  with the Automess gamma probe 6150AD-15 is comparable to the nominal attenuation of the 662~keV photons. For factors greater than 10 though, the effective dose attenuation is considerably lower than the nominal attenuation factor, since scattered photons with an energy smaller than 662~keV contribute substantially.

\begin{table}[htbp]
\centering
\scriptsize
\begin{threeparttable}
\begin{tabular}[b]{ l|r|r|r|}
\cline{3-4}
\multicolumn{2}{ l| }{} & \multicolumn{2}{ |c| }{\textbf{Measured data}}  \\ 
\hline
\multicolumn{1}{ |c|}{\textbf{Nominal}} & \multicolumn{1}{ |c|}{\textbf{Filter}}& \multicolumn{1}{ |c|}{\textbf{Dose}} & \multicolumn{1}{ |c|}{\textbf{Dose}}  \\ 
\multicolumn{1}{ |c|}{\textbf{Attenuation}} & \multicolumn{1}{ |c|}{\textbf{Combination}} & \multicolumn{1}{ |c|}{\textbf{Rate}} & \multicolumn{1}{ |c|}{\textbf{Attenuation}}  \\ 
\multicolumn{1}{ |l|}{\textbf{}} & \textbf{} & \textbf{[mGy/h]} & \textbf{}  \\ 
\hline
\multicolumn{1}{ |r|}{1} & A1~B1~C1 & 470.00 & -  \\
\multicolumn{1}{ |r|}{1.5} & A1~B2~C1 & 400.00 & 1.2 \\
\multicolumn{1}{ |r|}{2.2} & A1~B1~C2 & 211.00 & 2.2 \\
\multicolumn{1}{ |r|}{4.6} & A1~B1~C3 & 105.00 & 4.5 \\
\multicolumn{1}{ |r|}{10} & A2~B1~C1 & 55.00 &  8.8 \\
\multicolumn{1}{ |r|}{100} & A3~B1~C1 & 6.50 & 72.3 \\
\multicolumn{1}{ |r|}{100} & A1~B3~C1 & 6.20 & 75.8 \\
\multicolumn{1}{ |r|}{464} & A1~B3~C3 & 1.59 & 295.6  \\
\multicolumn{1}{ |r|}{4642} & A2~B3~C3 & 0.22 & 2156.0 \\
\multicolumn{1}{ |r|}{46415} & A3~B3~C3 & 0.05 & 9400.0 \\
\hline
\end{tabular}
\caption{Nominal attenuation factors (attenuation of the 662~keV photons) of some filter settings and measured effective attenuation in position D1 (x=0.65m, y=0.00m, z=1.10m).}
\label{table: attenuation measured}
\end{threeparttable}
\end{table}

Table~\ref{table: Simulated current attenuation} displays the nominal and simulated attenuation of the photon current in position U1. The values in the different energy ranges are expressed as ratio of unattenuated over attenuated current. As explained in Section~\ref{subsec:modification of current}, the nominal attenuation factor of the filters is the attenuation of the 662~keV photons. The attenuation for lower energy photons deviates from this factor. In general, photons with energies between 100~keV and 300~keV are attenuated to a larger extent than the attenuation factor implies, whereas photons with energies smaller than 100~keV and between 400~keV and 600~keV are attenuated less. Depending on the energy of the incoming photons, a different percentage of the photons Compton scatters or reacts via the photoelectric effect. Further, detectors to be tested at GIF++ are characterized by a variety of materials, shapes, used gases and conditions of operation. The effective attenuation with regard to different photon energies is thus also detector dependent and hard to predict. The attenuation filters for detector tests should therefore be chosen carefully, also considering the detector specific sensitivity.

\begin{table}[htbp]
\centering
\scriptsize
\begin{threeparttable}
\begin{tabular}[htbp]{ |l|r|r|r|r|r|r|r|r|r|}
\hline
\rot{90}{\shortstack[c]{Nominal\\\phantom{0}Attenuation\phantom{0}}}& \rot{90}{\shortstack[c]{0-100\\keV}} & \rot{90}{\shortstack[c]{100-200\\keV}} & \rot{90}{\shortstack[c]{200-300\\keV}} & \rot{90}{\shortstack[c]{300-400\\keV}}  & \rot{90}{\shortstack[c]{400-500\\keV}} & \rot{90}{\shortstack[c]{500-600\\keV}}
& \rot{90}{\shortstack[c]{600-662\\keV}} & \rot{90}{\shortstack[c]{0-662\\keV}} & \rot{90}{\shortstack[c]{661-662\\keV}}\\
\hline
\multicolumn{1}{ |l|}{1.0 } & 1.00 & 1.00  & 1.00  & 1.00  & 1.00  & 1.00  & 1.00  & 1.00  & 1.00  \\
\hline
\multicolumn{1}{ |l|}{1.5} & 0.74 & 4.11 & 2.47 & 1.41 & 1.15 & 1.18 & 1.41 & 1.42 & 1.45  \\
\multicolumn{1}{ |l|}{2.2} & 1.19 & 7.3 & 5.7 & 2.10 & 1.56 & 1.55 & 2.02 & 2.0 & 2.12  \\
\multicolumn{1}{ |l|}{3.2} & 1.63 & 10.4 & 8.6 & 2.89 & 1.94 & 1.97 & 2.87 & 2.8 & 3.10  \\
\multicolumn{1}{ |l|}{4.6} & 2.3 & 13.5 & 13.2 & 4.28 & 2.80 & 2.68 & 4.09 & 3.9 & 4.50  \\
\multicolumn{1}{ |l|}{6.8} & 3.3 & 18.8 & 18.5 & 5.9 & 3.65 & 3.50 & 5.78 & 5.4 & 6.54  \\
\multicolumn{1}{ |l|}{10} & 5.0 & 19.6 & 19.0 & 8.8 & 4.69 & 4.50 & 8.22 & 7.3 & 9.47  \\
\multicolumn{1}{ |l|}{14.7} & 6.3 & 32 & 30 & 11.8 & 6.2 & 5.85 & 11.8 & 10 & 13.8  \\
\multicolumn{1}{ |l|}{21.5} & 8.7 & 47 & 58 & 17.7 & 9.3 & 8.2 & 16.7 & 15 & 19.9  \\
\multicolumn{1}{ |l|}{31.6} & 12.6 & 62 & 84 & 24.1 & 12.0 & 11.1 & 24.1 & 20 & 29.5  \\
\hline
\end{tabular}
\caption[Nominal and simulated attenuation of photon current in position U1 (x=0.65m, y=0.00m, z=-1.10m). The values in the different energy ranges are expressed as ratio of unattenuated over attenuated current.]{Nominal and simulated attenuation of photon current in position U1. The values in the different energy ranges are expressed as ratio of unattenuated over attenuated current. The intensity of the unattenuated photon current in U1 can be found in Table~\ref{table: Simulated current}.}
\label{table: Simulated current attenuation}
\end{threeparttable}
\end{table}

\begin{figure}
\centering
\captionsetup[subfloat]{justification=centering}
\subfloat[Attenuation factors 1, 10 and 100\label{fig: Spectra attenuation factors}]{
\includegraphics[width=.49\textwidth]{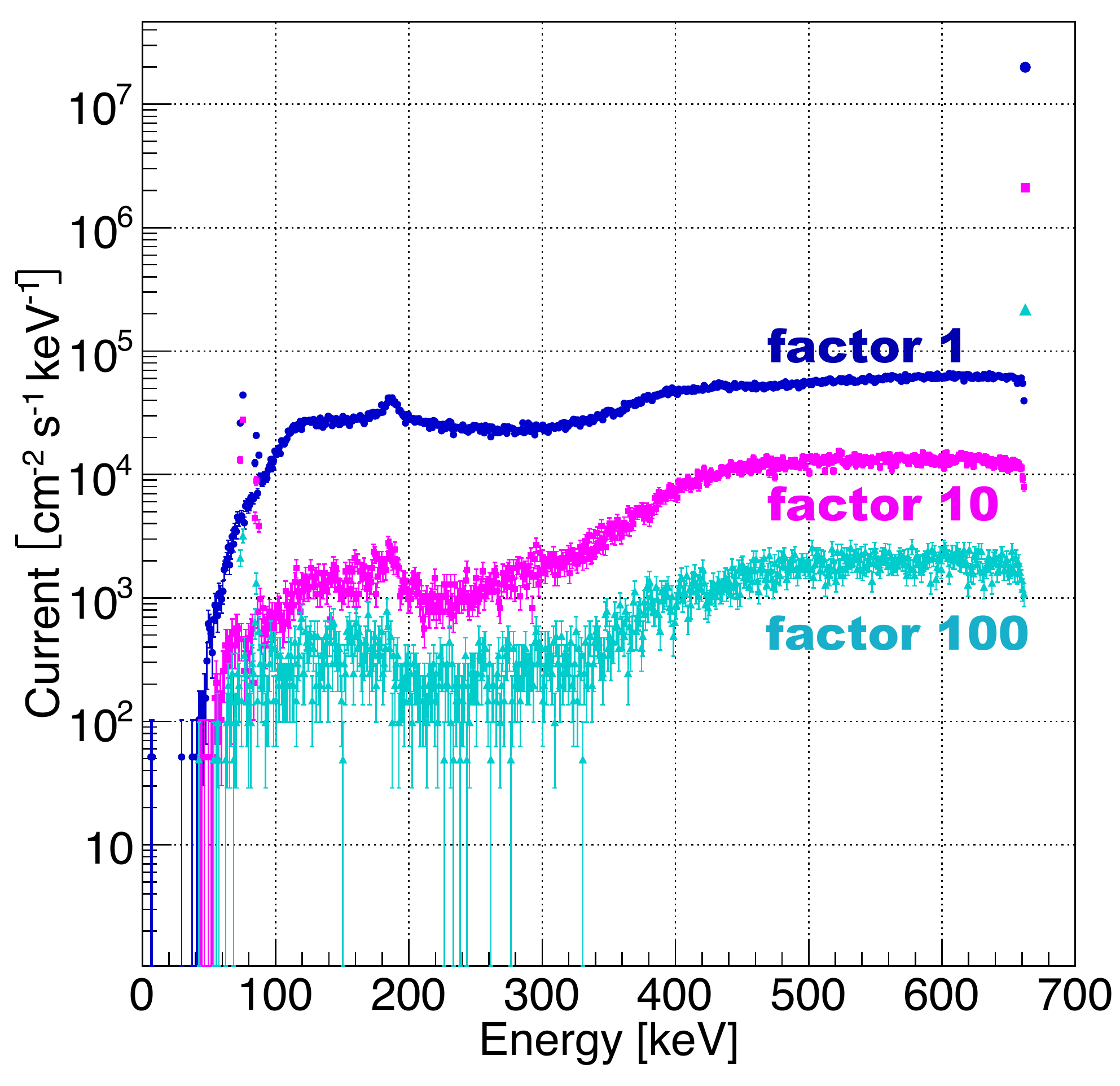}%
}
\subfloat[Crosstalk between D and U\label{fig: crosstalk}]{
\includegraphics[width=.49\textwidth]{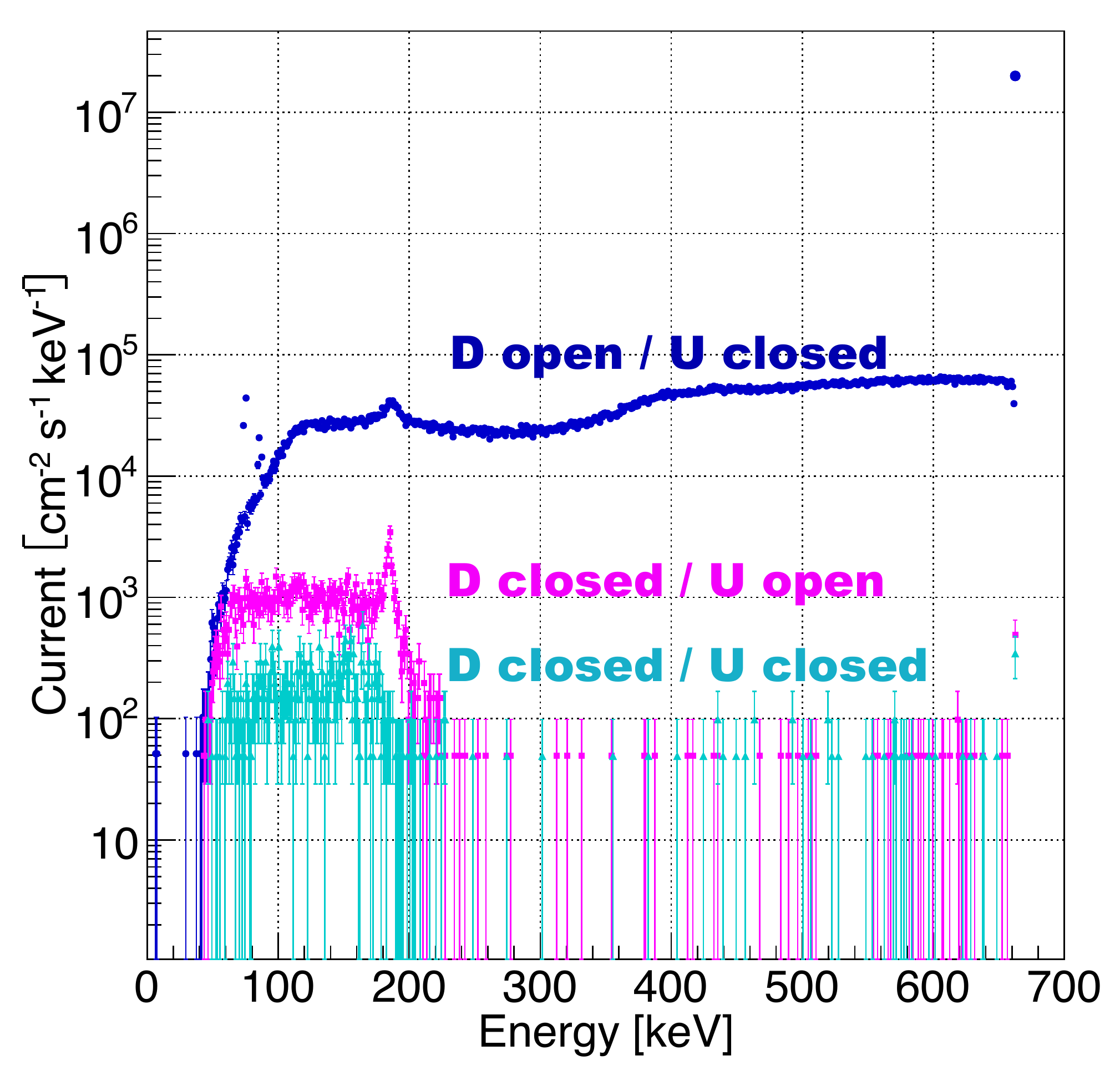}%
}
\caption{Simulated spectra (bin size 5~keV) in location D1: a) Downstream attenuation factors of 1, 10 and 100 with upstream side closed; b) Crosstalk between downstream and upstream side with D open/U closed, D closed/U open and D closed/U closed. When D is open, the narrow and very high main 662~keV peak can be seen in both figures at the far right.}
\label{fig: Attenuation factors}
\end{figure}

Figure~\ref{fig: Spectra attenuation factors} shows the spectra of the photon current in location D1 for the attenuation factors 1, 10 and 100. In addition to the narrow 662~keV peak, a broad low energy component is visible. In fact the photons arrive already at the attenuation filters with this low energy component. Before even reaching the angular correction filter, the source capsule and the irradiator collimator already cause an amount of scattering that broadens the spectrum. Scattering in the angular correction filters and the bunker (walls, floor and roof) adds further low energy contributions to the spectrum. As expected, the intensity of the photons with an energy of 662~keV is reduced by a factor of 10 or 100, in agreement with the nominal attenuation of the lead filters. Equivalent to the situation in position U1 described in Table~\ref{table: Simulated current attenuation}, the 400~keV to 600~keV photons are attenuated less, and the 100~keV to 300~keV photons are attenuated more than the nominal attenuation factor. Clearly visible in the spectra are the back-scatter peak at 184 keV and the characteristic $^{82}$Pb K-shell X-ray peaks around 80~keV, which cannot be resolved individually though, due to the energy binning. Both features are explained in more detail in Section~\ref{sec: albedo studies}.

Figure~\ref{fig: crosstalk} shows three spectra in location D1, one with the downstream filters fully open and the upstream filters fully closed, one with the downstream filters fully closed and the upstream filters fully open and the third one with both sides fully closed. The two spectra with D closed show that when opening U, the rate in D increases significantly. By Compton scattering multiple times, photons from the upstream opening manage to arrive in the downstream zone of the facility. The absolute intensity of this \textit{crosstalk} photon current is nevertheless small. Therefore the field in the measurement zone can be regarded as basically independent from the field in the opposing zone, unless a very large attenuation factor is used in the measurement zone.

\subsection{Simulations of degraded photons and albedo}
\label{sec: albedo studies}
\begin{figure}[htbp]
\centering
\captionsetup[subfloat]{justification=centering}
\subfloat[Degraded photons from the irradiator\label{fig: energy distribution degraded photons}]{
\includegraphics[width=.475\textwidth]{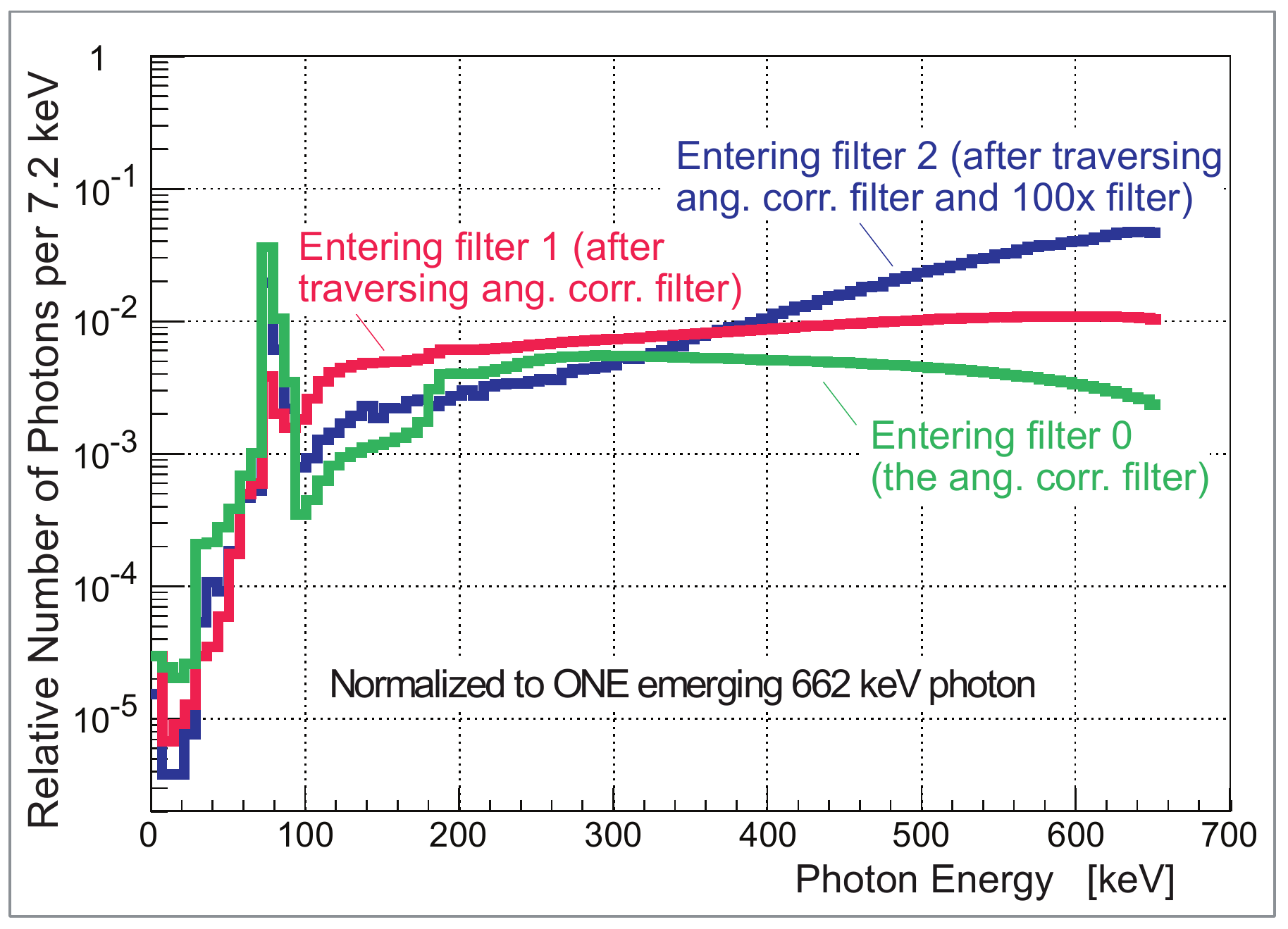}% 
}
\subfloat[Albedo photons from a wall\label{fig: energy distribution albedo}]{
\includegraphics[width=.505\textwidth]{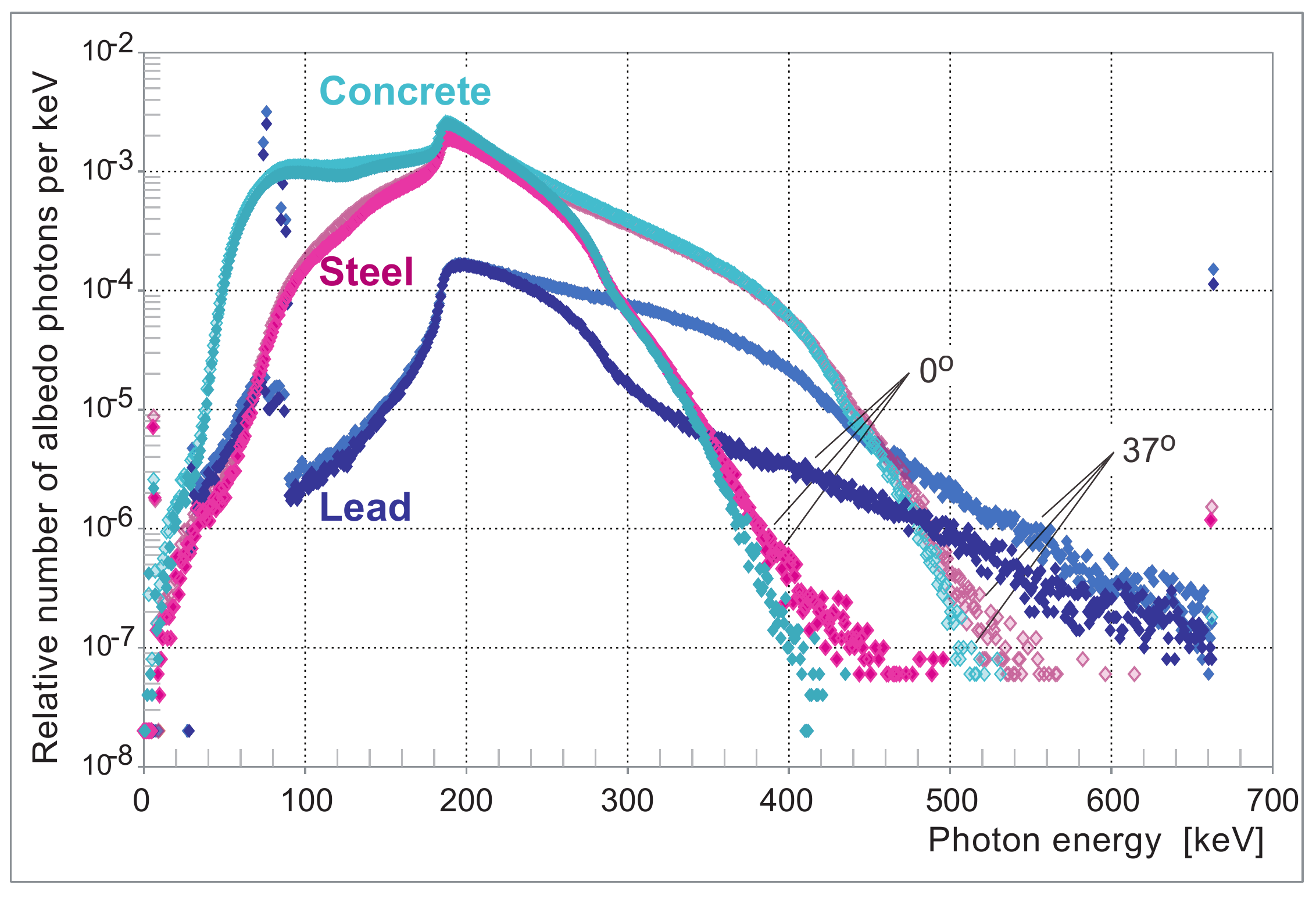}%
}
\caption{Calculated energy distribution of (a) degraded photons emerging from the irradiator, per emerging 662~keV photon, and (b) of albedo photons from 662~keV photons impinging on a wall, for three wall materials, per 662~keV incident photon.}
\label{fig: Degraded}
\end{figure}

The photon current is modified by the interaction of photons with the materials present in the GIF++ facility. The materials concerned are lead (irradiator, filters), steel (filters, floor), aluminium (filters) and concrete (surrounding bunker enclosure). For the available photon energies of 662~keV or below, the main processes involved are Compton scattering and photoelectric effect. Following the photoelectric effect in lead, there is a high probability that the atom de-excites by emitting fluorescent X-rays. To better understand the photon current modification, these processes were simulated for photons of different energies and angles of incidence. Through collisions with electrons of dense materials, photons may loose a fraction of their energy along their travel. Therefore, in addition to the monochromatic 662~keV photons from $^{137}$Cs, a low energy component of \textit{degraded} photons develops within the irradiator. This low energy component is already present when the photons emerge from the small thin-walled capsule containing the active material\footnote{The wall of the capsule is actually 1.75 mm thick, 1 mm thicker than assumed in the simulation.}. Interactions in the material subsequently traversed further modify the spectral shape of degraded photons, as illustrated in Figure~\ref{fig: energy distribution degraded photons} for photons emerging from the capsule, from the angular correction filter, and from an exemplary attenuation filter of nominal factor 100. The peak around 80~keV corresponds to the $K_\alpha$ fluorescence line expected for lead. The vertical scale is normalized to one emerging 662~keV photon. 
For large attenuation factors, a general trend is visible. In comparison with the nominal attenuation factor for the 662~keV photons, the attenuation is more pronounced below 400~keV, but less pronounced above 400~keV. The fraction of degraded photons is coarsely the same at all stages, that means the number of degraded photons is also reduced when the number of 662~keV photons is reduced by absorption filters. The degraded photons amount to about 46$\%$ of the photons that reach the angular correction filter. Behind the angular correction filter, 39$\%$ of the photons that emerge from the irradiator within its 37$^\circ$ by 37$^\circ$ aperture are degraded. Behind the filter with a nominal attenuation of 100, the final fraction of degraded photons reaches 56$\%$.

All photons emitted by the irradiator will interact with the surrounding material in the bunker and will finally be absorbed. The photons that are not absorbed when impinging on a wall, but manage to re-enter into the bunker, are called \textit{albedo} photons. They are expected to be mainly back-scattered photons from Compton scattering; for lead some fluorescence photons contribute to the albedo. The calculated energy distribution of albedo photons is shown in Figure~\ref{fig: energy distribution albedo} for 662~keV photons impinging with angles of 0$^\circ$ and 37$^\circ$ with respect to the surface normal on a wall of infinite thickness and made of concrete, steel or lead. The vertical scale is per photon incident on the wall.

While colliding with a nearly free electron, the lowest energy of the scattered photon corresponds to a scattering angle of 180$^\circ$. For 662~keV photons this leads, independent of the wall material, to a back-scattered photon of about 184~keV. Photons back-scattered at other angles will have somewhat higher energy. Inside the wall, before emerging again, the photon may suffer further Compton scatterings, which will lower the energy of the albedo photon. Also incident photons of energies lower than 662~keV will extend the albedo spectrum to lower energies. As visible in Figure~\ref{fig: energy distribution albedo}, the concrete and steel spectra are dominated by this Compton back-scatter peak, and it is also visible in the lead spectrum. The contribution from the photoelectric effect, approximately proportional to Z$^3$ (Z is the nuclear charge of the target material) and to about E$^{-3}$ (E is the energy of the incident photon), is therefore only relevant for lead in terms of probability and of energy. The characteristic K$\alpha$ and K$\beta$ X-ray fluorescence lines~\cite{X-ray} of 72 to 88~keV are prominent in the lead albedo spectra. For all three materials, the spectra at 0$^\circ$ and at 37$^\circ$ incidence are almost identical, except in the tail region above 250~keV. At 0$^\circ$ (37$^\circ$) incidence the fraction of albedo per incident 662~keV photon is 24.6$\%$ (29.5$\%$) for concrete, 14.7$\%$ (18.6$\%$) for steel and 1.9$\%$ (2.8$\%$) for lead.

\begin{figure}[htbp]
\centering
\captionsetup[subfloat]{justification=centering}
\subfloat[Back-scatter probability of photons\label{fig: albedo fraction}]{
\includegraphics[width=.511\textwidth]{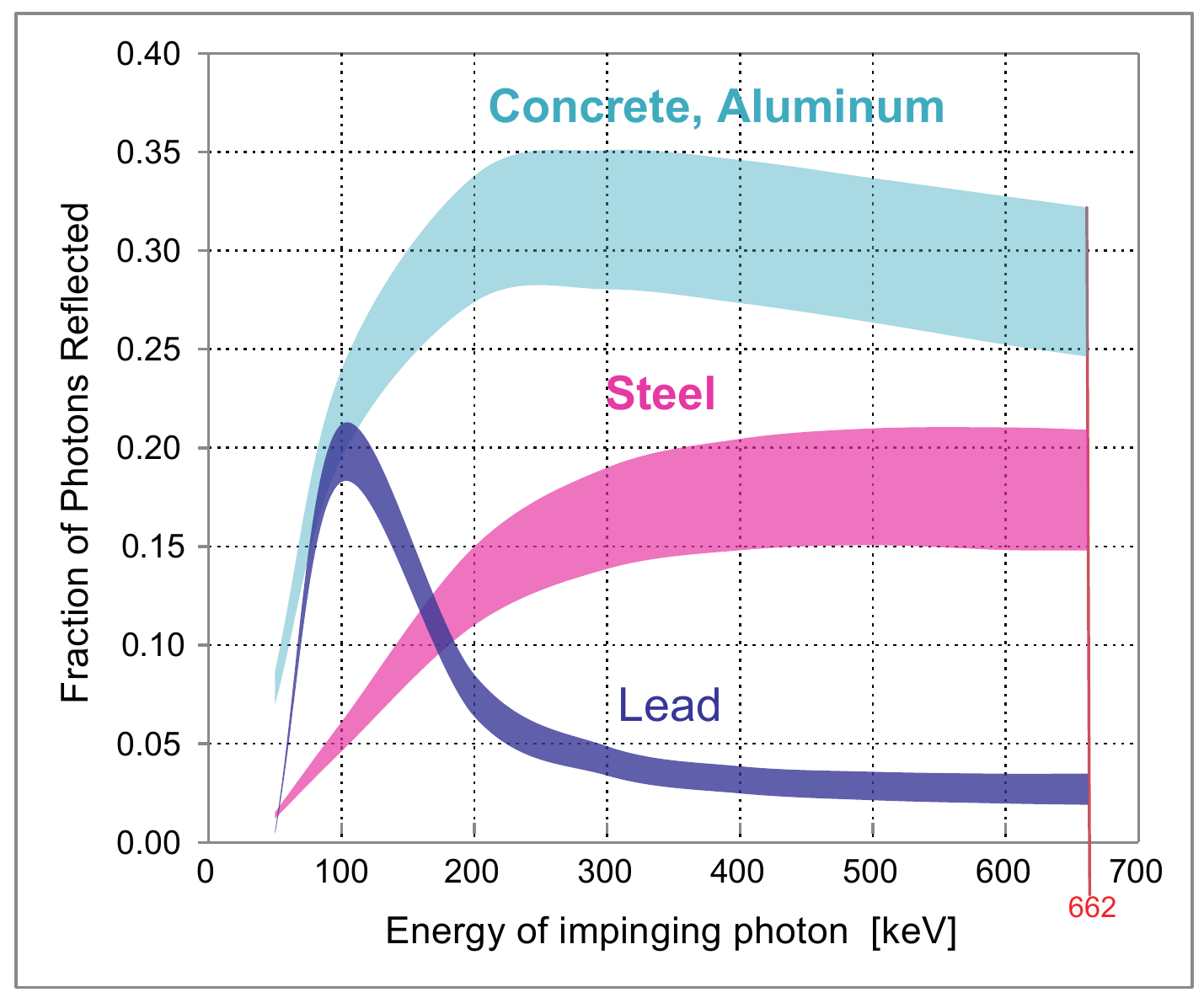}%
}
\subfloat[Penetration depth of reflected photons\label{fig: albedo depth}]{
\includegraphics[width=.48947\textwidth]{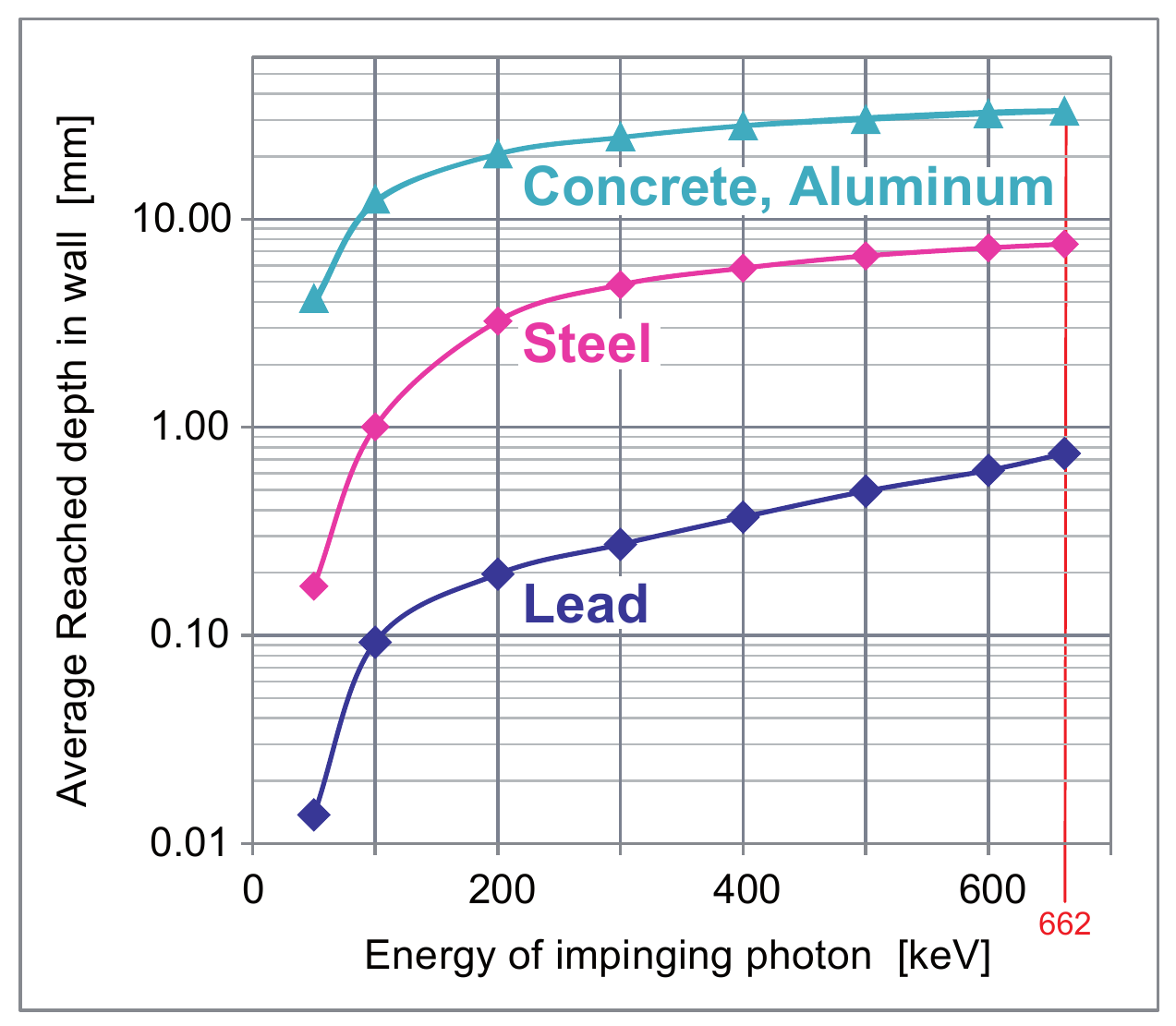}%
}
\caption{Back-scatter probability of incoming photons with incidence between 0$^\circ$ and 45$^\circ$ and penetration depth of reflected photons depending on material and incoming photon energy and angle. }
\label{fig: Albedo}
\end{figure}

To give an overview of the amount of albedo under different conditions relevant at GIF++, Figure~\ref{fig: albedo fraction} illustrates the dependence of albedo on the energy of the incident photon, for incident angles from 0$^\circ$ to 45$^\circ$. The lower(higher) border of the bands is for 0$^\circ$(45$^\circ$). The influence of the angle of incidence on the back-scatter probability is small. Down to about 250~keV the probability of \textit{reflecting} a photon is roughly the same as for 662~keV. In this main region of energies it is an order of magnitude lower for lead than for concrete. Below 250~keV the probability decreases, except for lead which shows a large increase related to the photoelectric effect.

It is interesting to see how deep a photon penetrates into a wall before emerging again. The calculated average depth reached by these photons, shown in Figure~\ref{fig: albedo depth}, is basically independent of the angle of incidence. It decreases slowly from the largest depth at 662~keV down to about 200~keV; the standard deviation has almost the same value as the average. With a wall thickness of twice this average depth the albedo production is thus saturated. Therefore with a 2 mm thick lead layer on walls, ceiling and floor, the amount of albedo photons could be reduced by one order of magnitude with respect to concrete. The large penetration depth in concrete, together with the angular distribution of the scattered photons inside the wall material has as consequence that the photons will emerge at a location, which may be different from the location of the impinging photon. The emission of albedo photons is isotropic in the azimuth angle with regard to the surface normal of the scattering surface. The polar emission angle is almost independent from the energy and the angle of incidence of the incoming photon from the source. At an incidence angle of 0$^\circ$ and 37$^\circ$, the average polar emission angle amounts to 50$^\circ$. At the wall surface the exit point of albedo from 662 keV photons is on average at about 3~cm distance from the impact point, with a 3.5~cm standard deviation. At 100~cm from the wall, the exit point is on average 160~cm away from the impact point, with 68$\%$ of the photons having an exit point that is less than 155~cm away. This shows that albedo is expected to be widespread in the bunker. 

The angular distribution of albedo photons, as they are originating at points distributed over the whole GIF++ bunker, is of course very different from the distribution for photons from the irradiator. The final distributions of photon rate, angle, energy in the GIF++ bunker will follow the principles described in this section, while reflecting the complexity of the actual geometry of the bunker and of all the installed equipment like detectors in the facility. Simulations providing a guideline for GIF++ as a whole are described in the following sections.

\subsection{Measurements and simulations of the dose rate}
\label{sec:dose}
\begin{figure}[htbp]
\centering
\captionsetup[subfloat]{justification=centering}
\subfloat[D open/U closed\label{fig: DS}]{
\includegraphics[width=.49\textwidth]{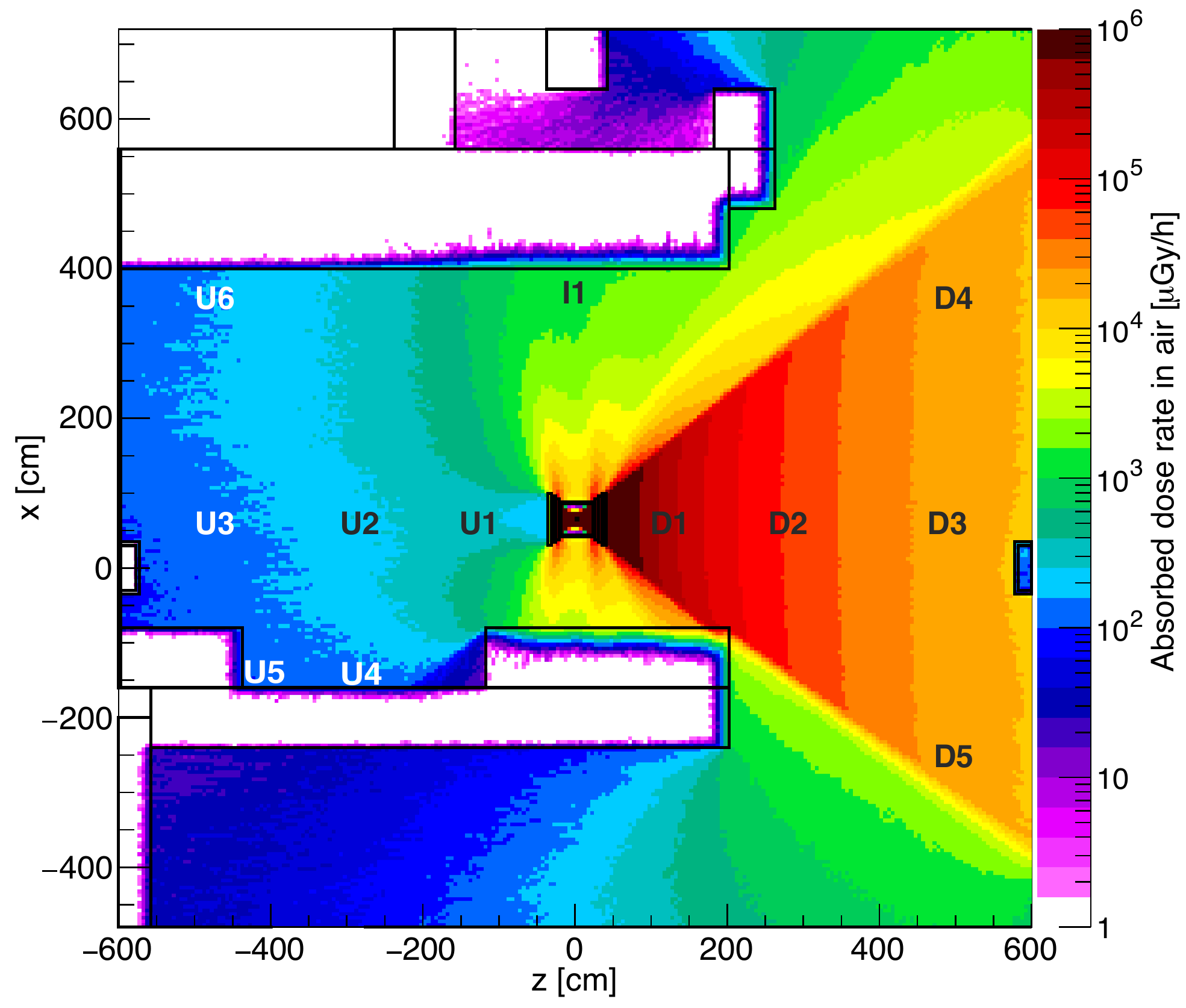}%
}
\subfloat[D closed/U open\label{fig: US}]{
\includegraphics[width=.49\textwidth]{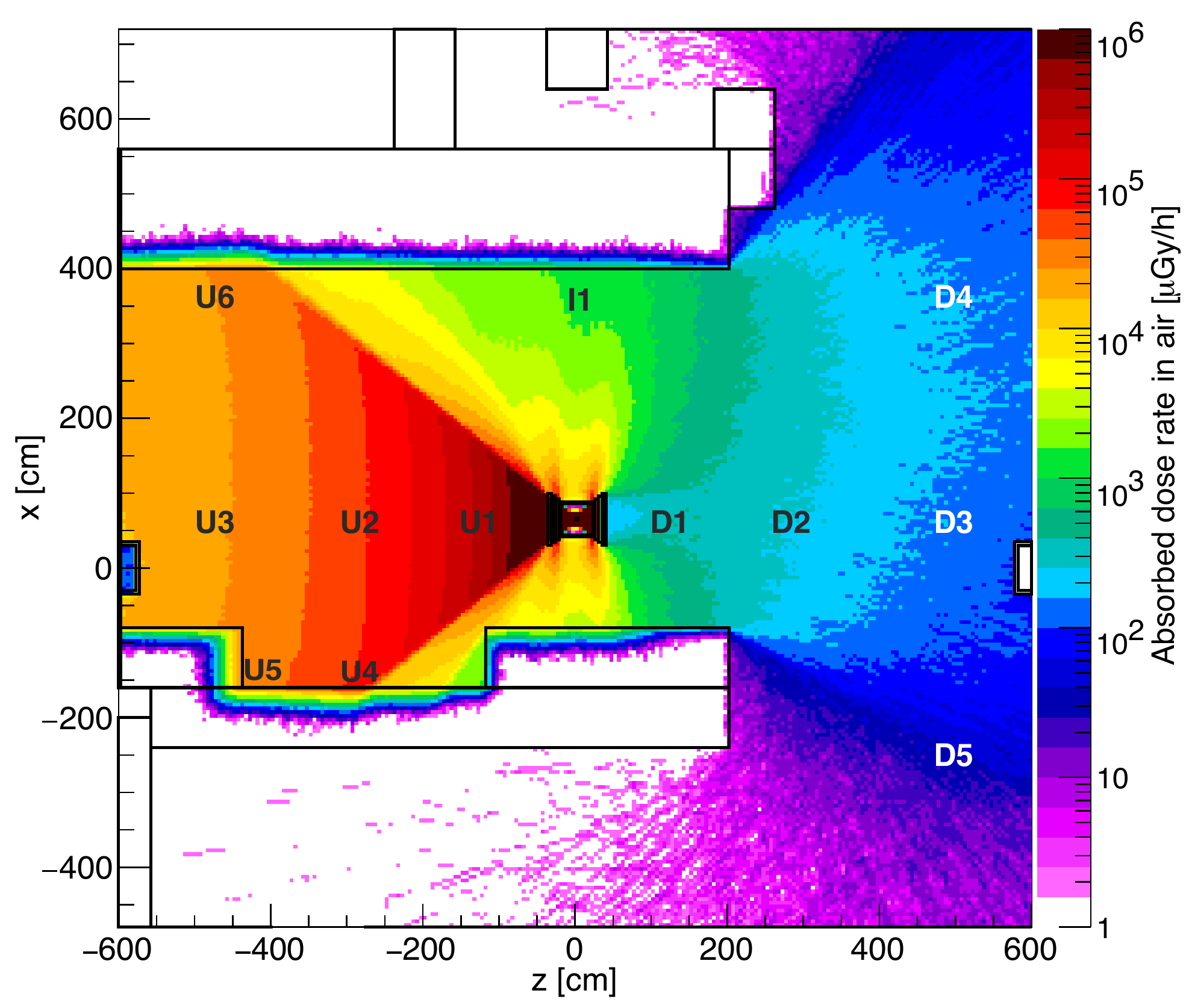}%
}
\caption{Absorbed dose rate in air [$\mu$Gy/h] in xz plane at y=0.0m of the facility. }
\label{fig: Dose}
\end{figure}

Figure~\ref{fig: Dose} shows the simulated absorbed dose rate in air in the xz plane of the GIF++ bunker. The measurement locations are marked in black. In Figure~\ref{fig: DS} the irradiator was fully open downstream and fully closed upstream, whereas in Figure~\ref{fig: US} the irradiator was fully closed downstream and fully open upstream. A fully closed irradiator outlet means an attenuation factor of 46415. The ratio of the doses in U1 and D1 reflects this attenuation in both figures. They demonstrate that the photons scatter significantly from the bunker walls, floor and roof. A quiet spot in the facility does not exist.\footnote{The dose contribution from photons passing through the irradiator shielding is negligible.} Nevertheless, the two figures confirm that the downstream and upstream zones of the GIF++ facility are basically independent. The intensity of the scattered photons depends on the exact position in the facility. A position closer to the walls like U6 suffers more from albedo than position D5, but both positions see about the same amount of direct photons. Furthermore the figures show that the angular correction filters do not only create a uniform current of 662~keV photons over the xy-planes within the irradiation regions, but also lead to a dose distribution that almost exclusively depends on z.

\begin{figure}[htbp]
\centering
\captionsetup[subfloat]{justification=centering}
\subfloat[RADMON measurement positions\label{fig: RADMON positions}]{
\includegraphics[width=.446229\textwidth]{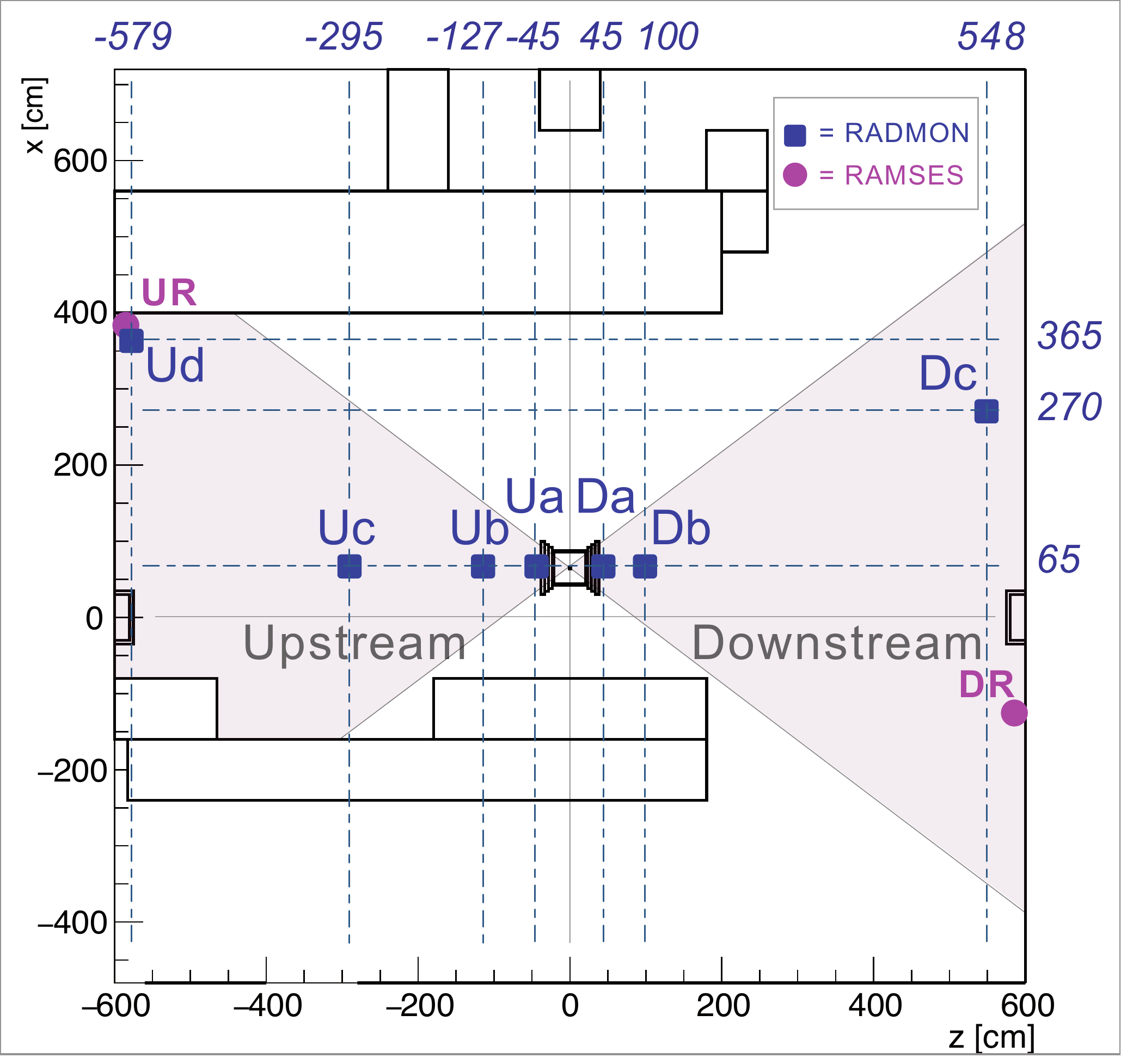}%
}
\subfloat[Measured and simulated data\label{fig: RADMON fit}]{
\includegraphics[width=.57347\textwidth]{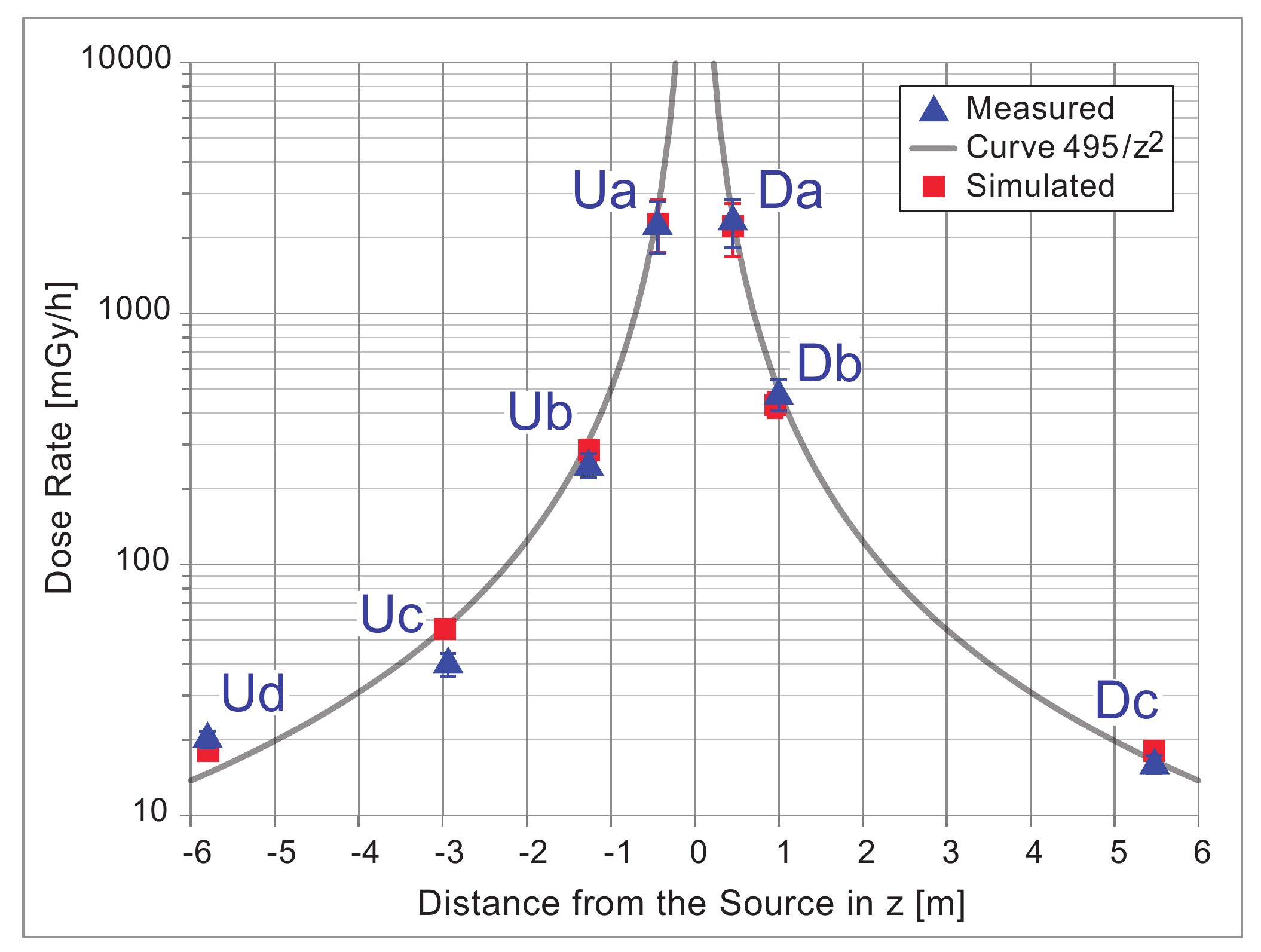}%
}
\caption{RADMON measurements of absorbed dose [mGy/h]. }
\label{fig: RADMON measurements}
\end{figure}

Measurements, as shown in Table~\ref{table: dose}, are compared with the simulations for the locations indicated in Figures~\ref{fig: Dose} and~\ref{fig: RADMON measurements}. For the simulations, in addition to the statistical error, a position uncertainty of $\pm$~5~cm was used, as taken from the estimated position uncertainty of the measurement probes. 

Given the Automess 6150AD-15 probe's allowed linearity deviation of $\pm$~10$\%$ and combined directional and energy deviation of $\pm$~40$\%$, a total dose measurement uncertainty of $\pm$~40$\%$ was assumed. The simulated dose rates are always larger than the measured dose rates with the Automess 6150AD-15 probe. In locations close to the irradiator, like the locations D1 and U1, where the majority of the dose is caused by 662~keV photons, measurement and simulation agree to about 15$\%$. Even at location I1, outside of the irradiation area, measurement and simulation agree within 40$\%$. Within the irradiation area, in locations with the same absolute value of the z-coordinate but different x coordinate, the measured dose rates vary less than 20$\%$. For the positions U3, U3a and U3b at different y coordinates, the variation is within 10$\%$. Dose rate for positions downstream and upstream, on the axis of the source and with the same distance from the source, agree within 10$\%$. This proves that the downstream and upstream zones of the facility have very similar radiation fields. 

The RAMSES measurement took place in positions at the upstream and downstream end wall of the bunker. The value in UR agrees well with the Automess measurement in U6, whereas the RAMSES measurement in DR lies between the Automess measurement and the simulated value.

\begin{table}
\scriptsize
\centering
\begin{threeparttable}
\begin{tabular}[htbp]{ |c|c|c|c|c|c|c|c|}
\cline{6-8}
\multicolumn{5}{ l| }{} & \multicolumn{3}{ |c| }{\textbf{Downstream open, Upstream open}} \\ 
\hline
\rot{90}{\shortstack[c]{\textbf{Position}}} & 
\rot{90}{\shortstack[c]{\textbf{x}\\\textbf{[m]}}}  &
\rot{90}{\shortstack[c]{\textbf{y}\\\textbf{[m]}}} & 
\rot{90}{\shortstack[c]{\textbf{z}\\\textbf{[m]}}} & 
\rot{90}{\shortstack[c]{\textbf{Detector}}} &
\multicolumn{1}{ |c|}{\rot{90}{\shortstack[c]{\textbf{\phantom{0}Measured dose rate\phantom{0}}\\\textbf{mGy/h}}}} &
\multicolumn{1}{ |c|}{\rot{90}{\shortstack[c]{\textbf{\phantom{0}Simulated dose rate\phantom{0}}\\\textbf{mGy/h}}}} & 
\multicolumn{1}{ |c|}{\rot{90}{\shortstack[c]{\textbf{\phantom{0}Measured dose rate\phantom{0}}\\\rule{2.8 cm}{0.3pt} \\ \textbf{\phantom{0}Simulated dose rate\phantom{0}} \\\textbf{[\%]}}}}\\
\hline
\multicolumn{8}{ |l|}{March 2015}\\
\hline
\multicolumn{1}{ |l|}{I1} & 3.65 & 0.0 & 0.0 &  6150AD-15 & 1.3(7) & 2.2(2)   & 57	\\
\hline
\multicolumn{1}{ |l|}{D1} & 0.65 & 0.0 & 1.1 & 6150AD-15 & -      & 410(42)  &  -    	\\
\multicolumn{1}{ |l|}{D2} & 0.65 & 0.0 & 2.9 & 6150AD-15 & -      & 56(3)    &  -    	\\
\multicolumn{1}{ |l|}{D3} & 0.65 & 0.0 & 4.9 & 6150AD-15 & 17(7)  & 22(1)    &  77	\\
\multicolumn{1}{ |l|}{D4} & 3.65 & 0.0 & 4.9 & 6150AD-15 & 15(6)  & 22(1)    & 69	\\
\multicolumn{1}{ |l|}{D5} & -2.35 & 0.0 & 4.9 & 6150AD-15 & 13(5)  & 21(1)    & 60	\\
\multicolumn{1}{ |l|}{DR} & -1.26 & 0.09 & 5.86 & RAMSES             & 12(5)	 & 15(1)	    & 76	\\
\hline
\multicolumn{1}{ |l|}{U1} & 0.65 & 0.0 & -1.1 & 6150AD-15 & 379(152)& 412(43)  & 92	\\
\multicolumn{1}{ |l|}{U2} & 0.65 & 0.0 & -2.9 & 6150AD-15 & 45(18)  & 55(3)    & 81	\\
\multicolumn{1}{ |l|}{U3} & 0.65 & 0.0 & -4.9 & 6150AD-15 & 17(7)  & 22(1)    & 79	\\
\multicolumn{1}{ |l|}{U3a}& 0.65 & -0.7 & -4.9 & 6150AD-15 & 20(8) & 23(1)     & 88	 \\
\multicolumn{1}{ |l|}{U3b}& 0.65 & -1.5 & -4.9 & 6150AD-15 & 18(7) & 23(1)     & 76 \\
\multicolumn{1}{ |l|}{U4} & -1.35 & 0.0 & -2.9 & 6150AD-15 & 43(17)  & 63(4)    & 68	\\
\multicolumn{1}{ |l|}{U5} & -1.35 & 0.0 & -4.2 & 6150AD-15 & 25(10)  & 32(1)    & 77	\\
\multicolumn{1}{ |l|}{U6} & 3.65 & 0.0 & -4.9 & 6150AD-15 & 16(6)  & 24(1)    & 67	\\
\multicolumn{1}{ |l|}{UR} & 3.8 & 0.05 & -5.90 & RAMSES             & 11(4)	& 17(1)	    & 66	\\
\hline
\multicolumn{8}{ |l|}{December 2015}\\
\hline
\multicolumn{1}{ |l|}{Da} & 0.65 & 0.0 & 0.45 & RADMON             & 2330(559) & 2213(521)	&	105		\\
\multicolumn{1}{ |l|}{Db} & 0.65 & 0.0 & 1.00 & RADMON             & 470(60) & 440(47)	&	107		\\
\multicolumn{1}{ |l|}{Dc} & 2.70 & 1.0 & 5.48 & RADMON             &	 16(1)	& 18(1)		&	89		\\
\hline
\multicolumn{8}{ |l|}{March 2016}\\
\hline
\multicolumn{1}{ |l|}{Ua} & 0.65 & 0.0 & -0.45 & RADMON             &	 2251(557)	& 2274(536)	&	99		\\
\multicolumn{1}{ |l|}{Ub} & 0.65 & 0.0 & -1.27 & RADMON             &	 249(30)		& 283(25)	&	88		\\
\multicolumn{1}{ |l|}{Uc} & 0.65 & 0.07 & -2.95 & RADMON             &	 40(4)		& 55(2)		&	73		\\
\multicolumn{1}{ |l|}{Ud} & 3.65 & 0.13 & -5.79 & RADMON             &  20(2)		& 18(1)		&	111		\\
\hline
\end{tabular}
\caption{Measured and simulated dose rate at the measurement locations in March 2015 (I1 to UR), December 2015 (Da to Dc) and March 2016 (Ua to Ud). Values in parentheses are the estimated uncertainties.}
\label{table: dose}
\end{threeparttable}
\end{table}	

For the RADMON measurement locations in Figure~\ref{fig: RADMON positions}, in addition to the $\pm$~5~cm position uncertainty, a precision uncertainty of $\pm$7.5~$\%$ has been estimated. Figure~\ref{fig: RADMON fit} shows a very good agreement ($\le$5$\%$ difference) between measurement and simulation for positions close to the source. The difference between measurement and simulations increases with the distance from the source, but stays below 12$\%$ for all positions with the exception of Uc. The deviation in Uc might be explained with the presence of other experimental equipment in the zone during the measurements. The GIF++ source, due to the angular correction filters, is expected to provide a dose rate dependence mainly on z instead of on r like a point source. Consequently, the comparison in Figure~\ref{fig: RADMON fit} is plotted as function of z. It shows that the power law is indeed a good approximation for the intensity distribution along the z-axis, especially close to the source.

To summarize, all measured values show within +11$\%$ and -43$\%$ agreement with the simulations. 

\subsection{Measurements and simulations of the photon current}
\label{sec: gamma current}
\begin{figure}[htbp]
  \centering
  \includegraphics[width=0.65\textwidth]{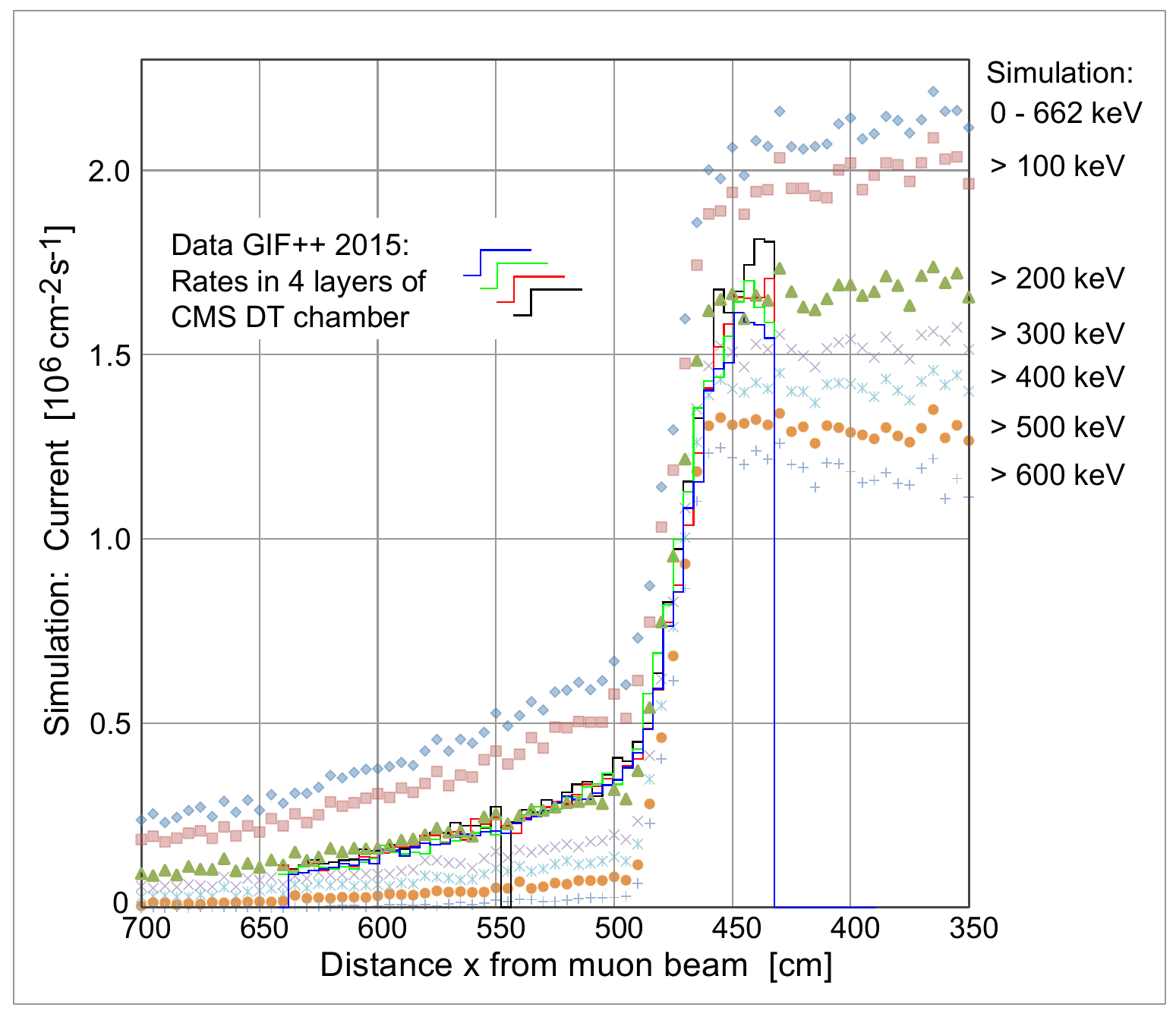}%
  \caption{Comparison of simulated photon current in air with the occupancy measured in drift cells of a CMS Drift Tube Chamber. The chamber was placed at the border of the irradiation cone. The steep drop of photon current outside the irradiation region is well visible in simulation and in data.}
   \label{fig: DT}
\end{figure}

As described in Section~\ref{subsec:modification of current}, the GIF++ irradiator and filter system is designed to produce a homogeneous radiation field of $\pm$~37$^\circ$, with a sharp decrease at the edges. In August 2015, a  Drift Tube (DT) chamber\footnote{The Drift Tube chamber consists essentially of 1.5 mm thick Aluminium plates. The 11.5 mm space between them is filled with Ar/CO$_2$ 85/15.} from the experiment CMS~\cite{DT}, was placed at 4.9~m distance from the source and at the edge of the 37$^\circ$ irradiation cone. The four solid lines of Figure~\ref{fig: DT} show the occupancy measured in four layers of 2.5~m long vertical drift cells of the 2~m wide drift chamber. The profiles measured by the four independent layers of 50 drift cells each look identical. Scaling the measured rates permits to compare the shape of the profile with simulation curves and it shows a good agreement for photon energies of at least 150~keV, suggesting that very low energy photons do not contribute significantly to the measured signal.This example demonstrates the capability of the simulation to also provide a description outside the borders of the irradiation cone. The width of this transition region is mainly related to the width of the source and its distance to the border of the collimator.

\begin{figure}[htbp]
\centering
\captionsetup[subfloat]{justification=centering}
\subfloat[Locations D1, D2, D3\label{fig: spectrum D1_3}]{
\includegraphics[width=.49\textwidth]{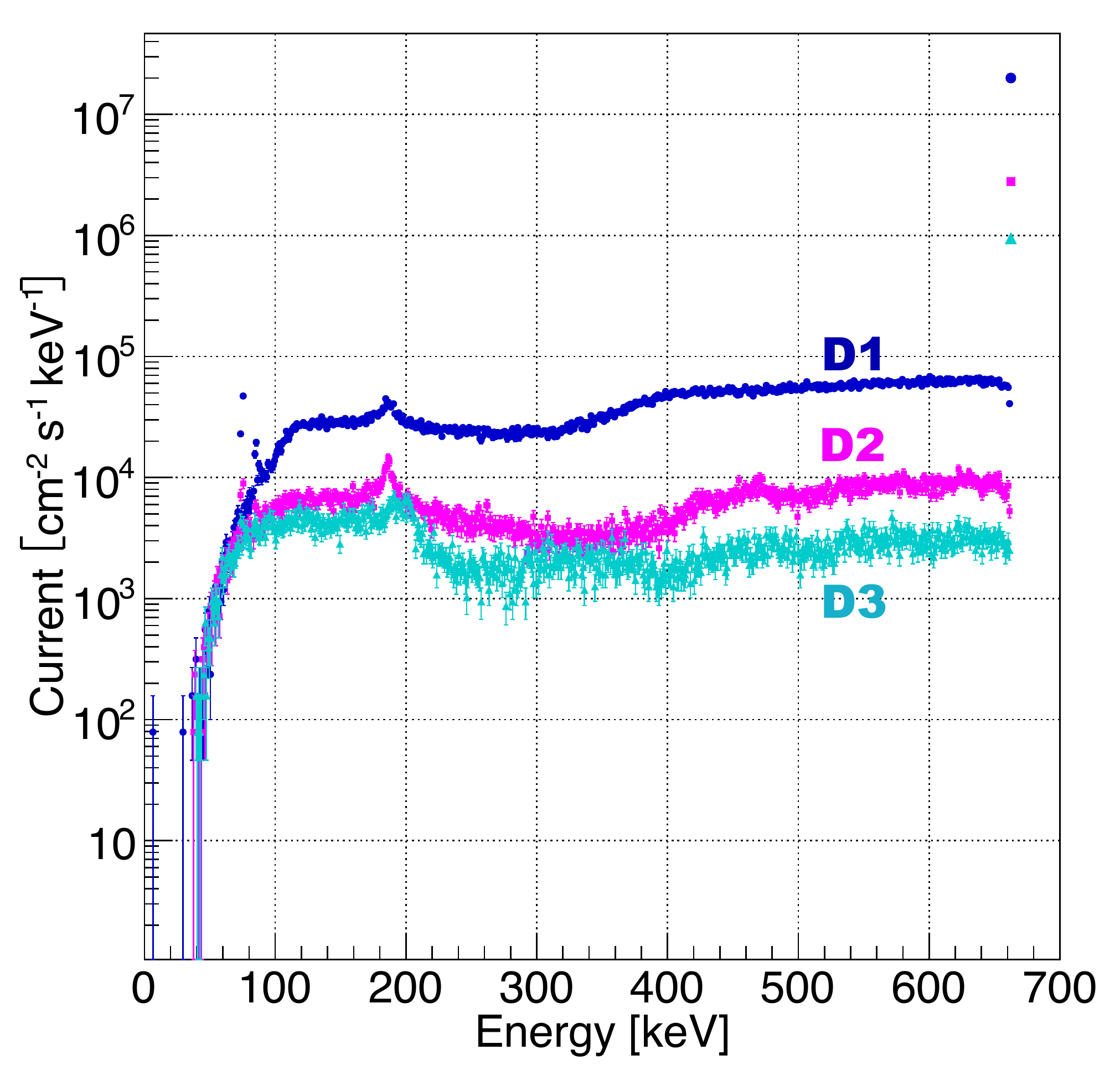}%
}
\subfloat[Locations U1, U2, U3\label{fig: spectrum U1_U3}]{
\includegraphics[width=.49\textwidth]{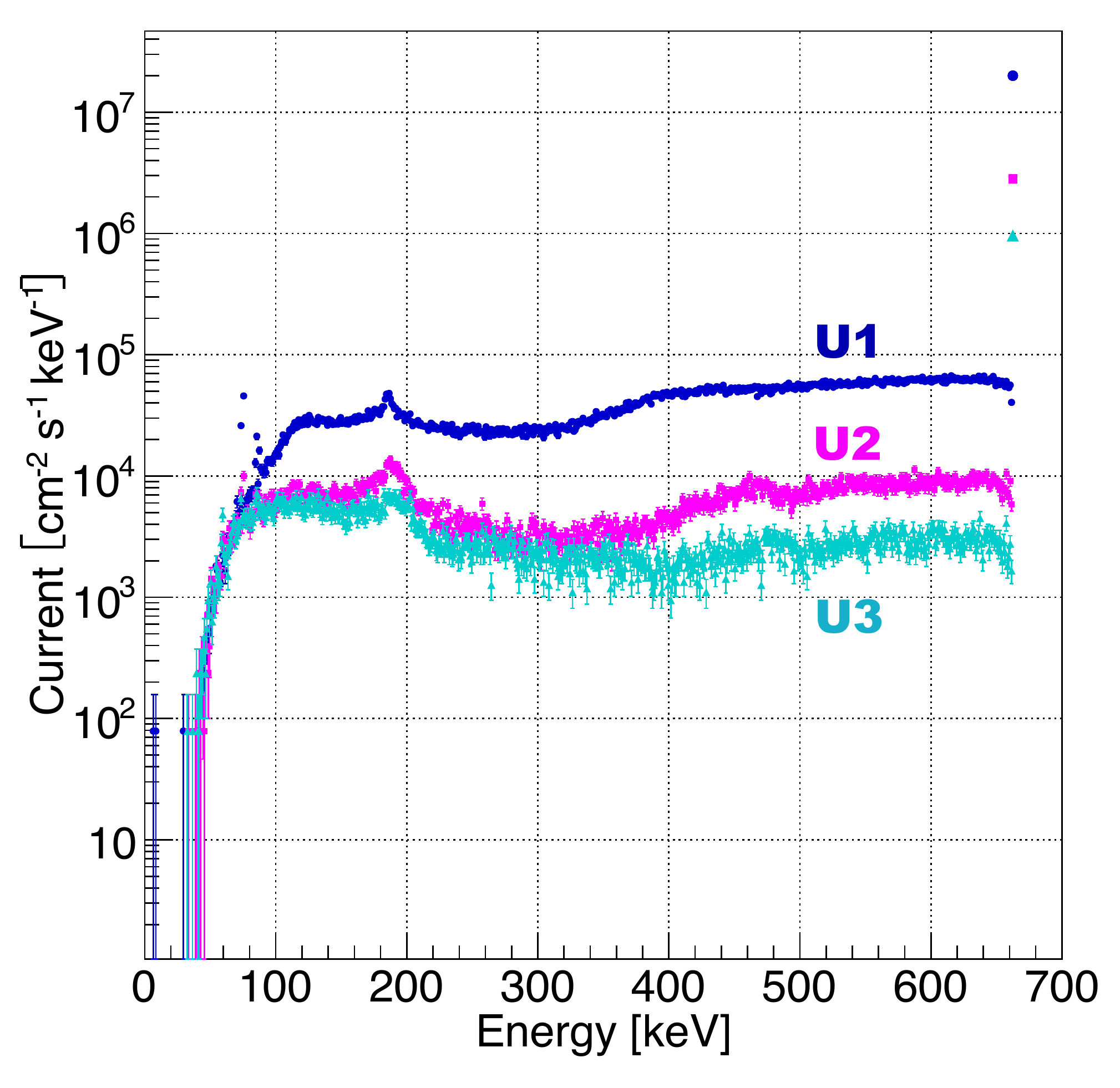}%
}
\\
\subfloat[Locations D4, D5, I1\label{fig: spectrum D4_I1}]{
\includegraphics[width=.49\textwidth]{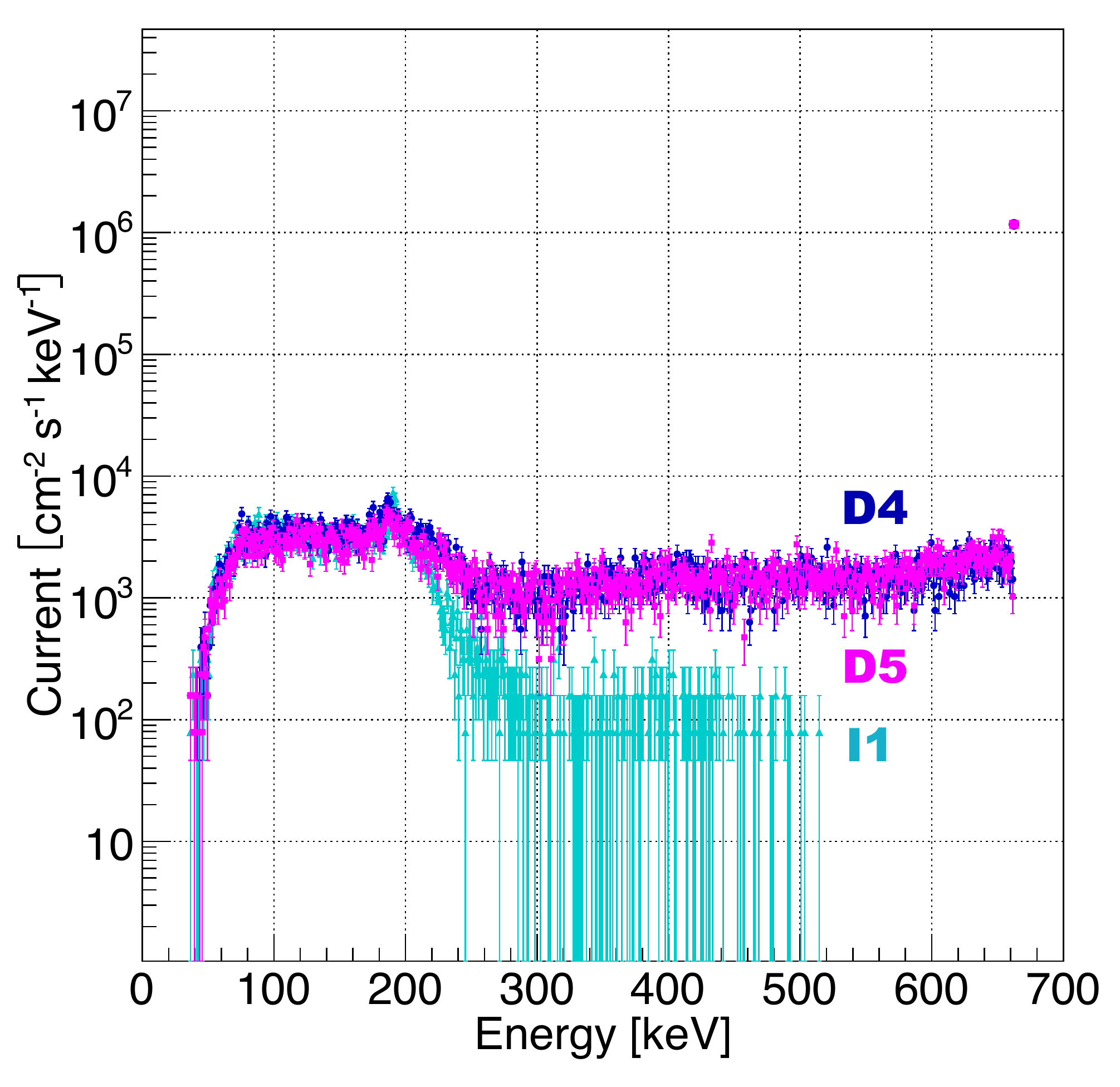}%
}
\subfloat[Locations U4, U5, U6\label{fig: spectrum U4_U6}]{
\includegraphics[width=.49\textwidth]{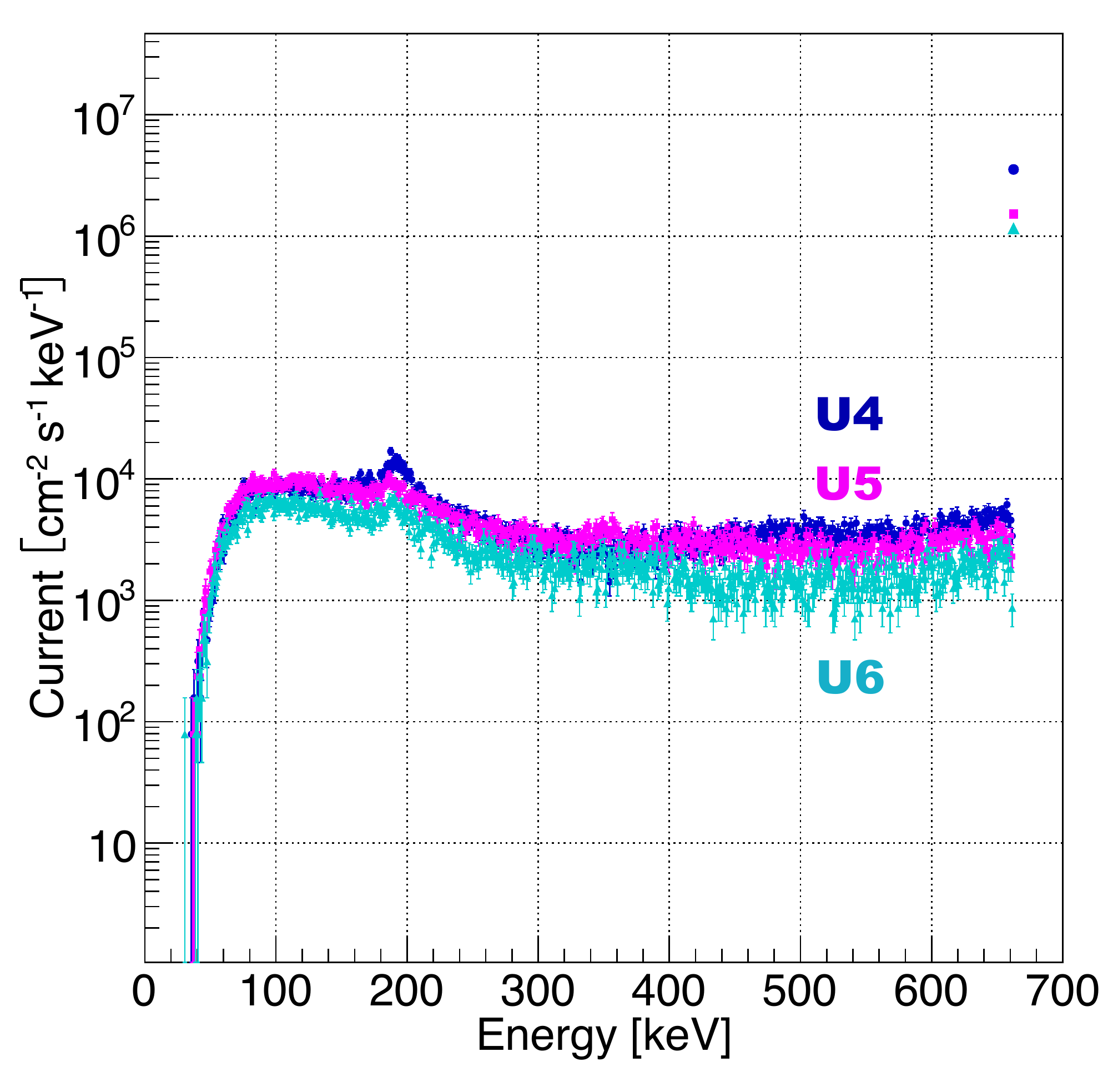}%
}
\caption{Simulated spectra (bin size 5~keV) in different locations with irradiator fully open downstream and upstream. The narrow and very high main 662~keV peak can be seen in all figures at the far right. In Figure (c), the 662~keV peaks for the locations D4 and D5 overlap, whereas for I1 the spectrum ends at about 500~keV.}
\label{fig: Spectra}
\end{figure}

To study the energy composition of the radiation field in GIF++, spectra of the photon current were simulated for the 14 measurement locations of March 2015. The spectra in Figure~\ref{fig: Spectra} confirm the results presented in Sections~\ref{sec: albedo studies} and~\ref{sec:dose}. Spectra in positions with the same z coordinate are basically identical. As expected, the spectrum in I1 does not contain any 662~keV photons, but consists of photons that scattered multiple times. In all other spectra the narrow and very high main 662~keV $^{137}$Cs peak is clearly visible at the far right. The concrete and steel back-scatter peak~(Section~\ref{sec: albedo studies}) is present in all spectra. In the spectra of positions U1 and D1, that are close to the irradiator and the filter systems, the characteristic $^{82}$Pb K-shell X-rays are present.

\begin{figure}[htbp]
\centering
\captionsetup[subfloat]{justification=centering}
\subfloat[Total current\label{fig: Current_DS_US_x}]{
\includegraphics[width=.49\textwidth]{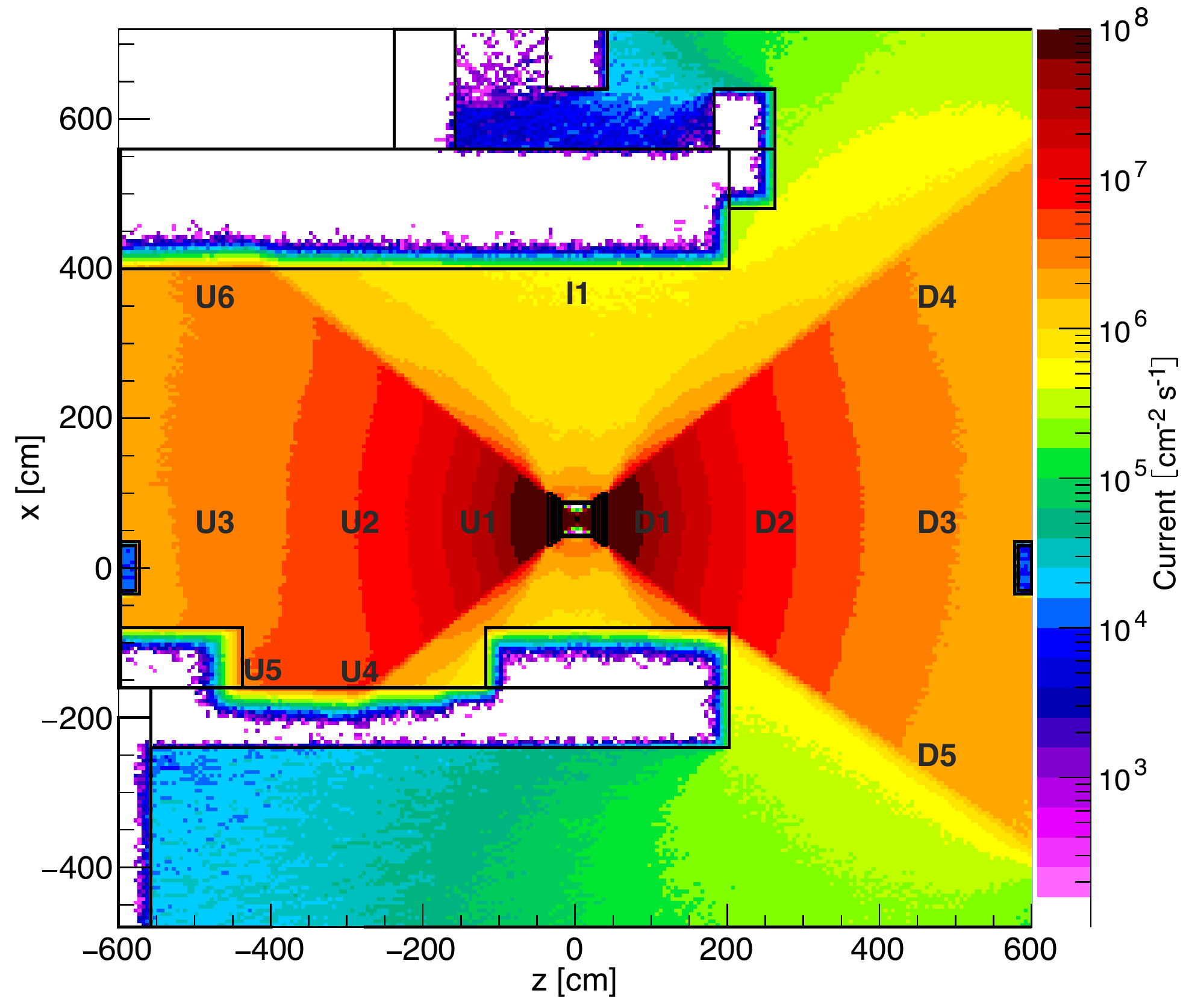}%
}
\subfloat[600~keV $<$ E $\le$ 662~keV \label{fig: Current_DS_US_x_600_667}]{
\includegraphics[width=.49\textwidth]{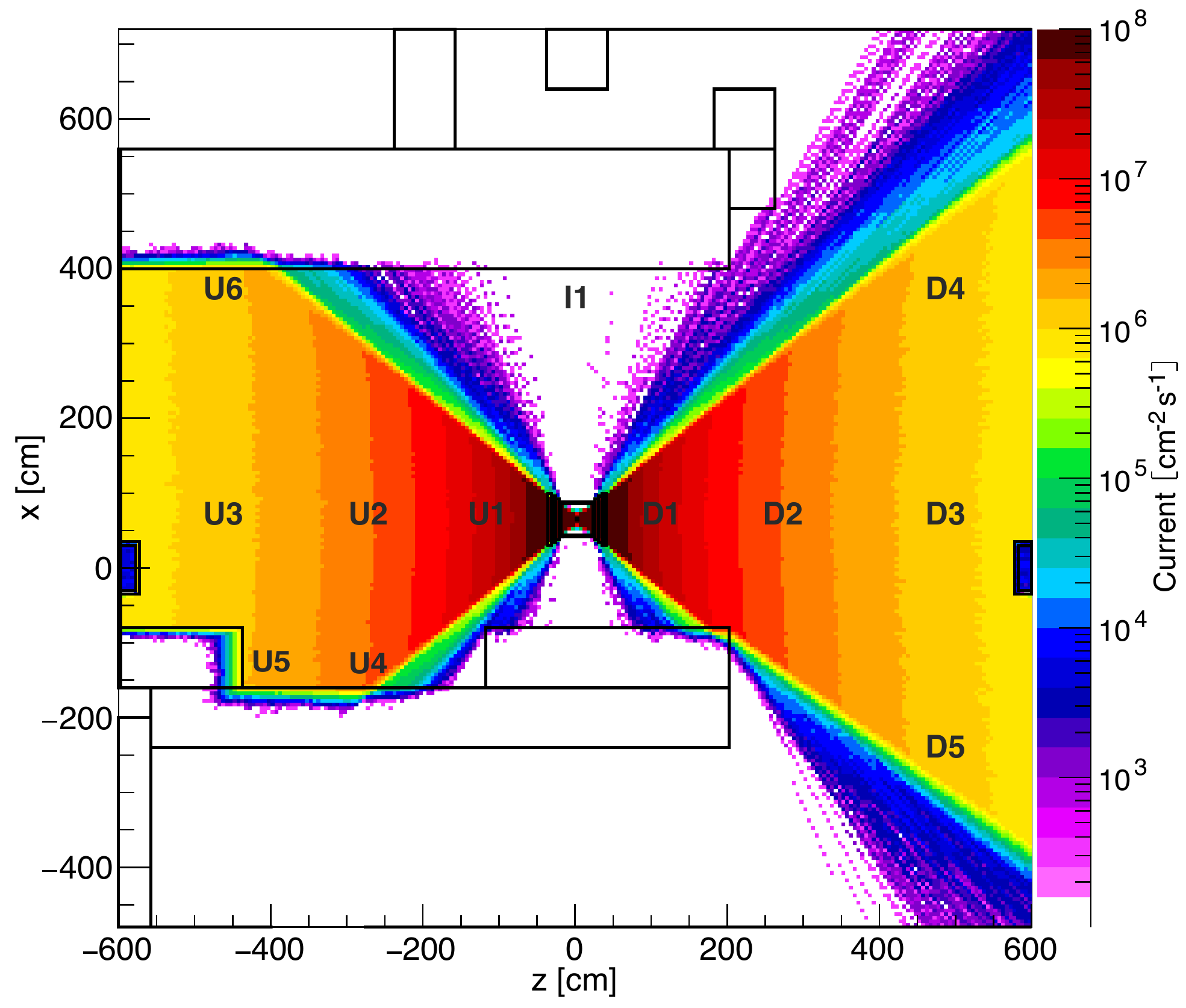}%
}
\\
\subfloat[500~keV $<$ E $\le$ 600~keV \label{fig: Current_DS_US_x_500_600}]{
\includegraphics[width=.49\textwidth]{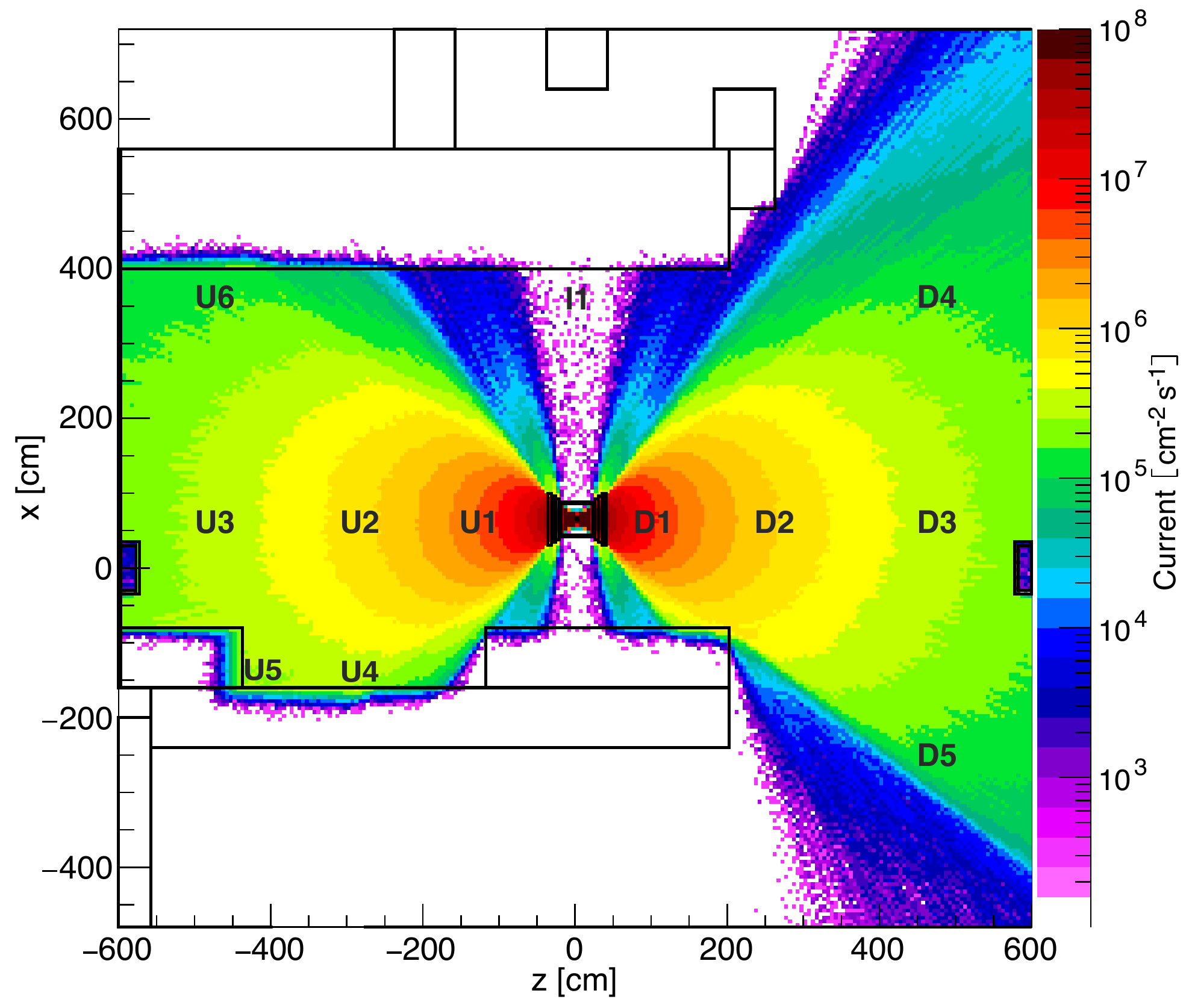}%
}
\subfloat[400~keV $<$ E $\le$ 500~keV \label{fig: Current_DS_US_x_400_500}]{
\includegraphics[width=.49\textwidth]{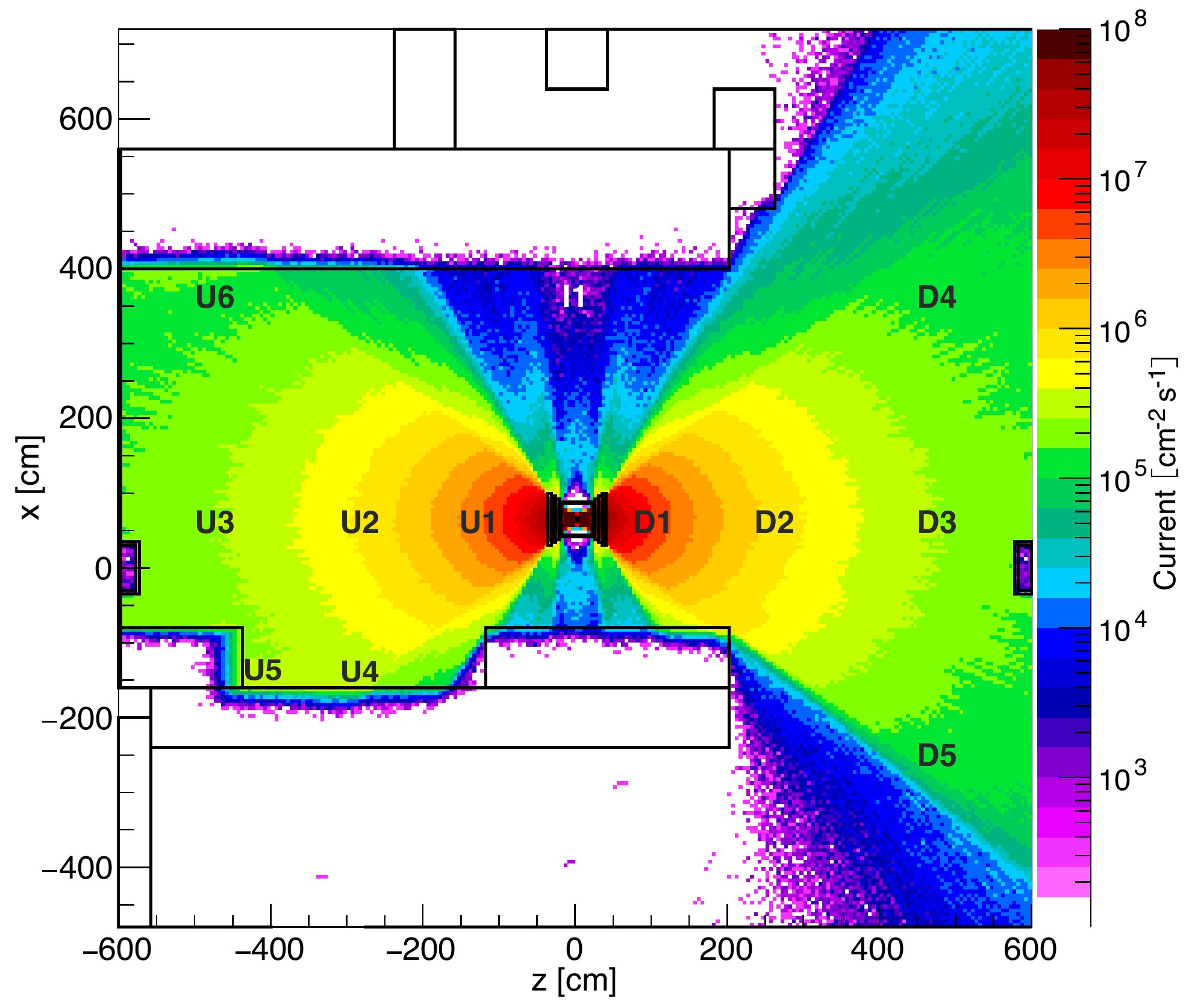}%
}
\caption{Simulated unattenuated current of photons in xz plane at y=0 m. The a) total current and the separate contributions of photons between b) 600~keV and 662~keV, c) 500~keV and 600~keV and d) 400~keV and 500~keV are shown. The largest total current amounts to 5$\times$10$^{7}$ photons/(cm$^2$ s).}
\label{fig: Simulated current 1}
\end{figure}

\begin{figure}[htbp]
\centering
\captionsetup[subfloat]{justification=centering}
\subfloat[300~keV $<$ E $\le$ 400~keV \label{fig: Current_DS_US_x_300_400}]{
\includegraphics[width=.49\textwidth]{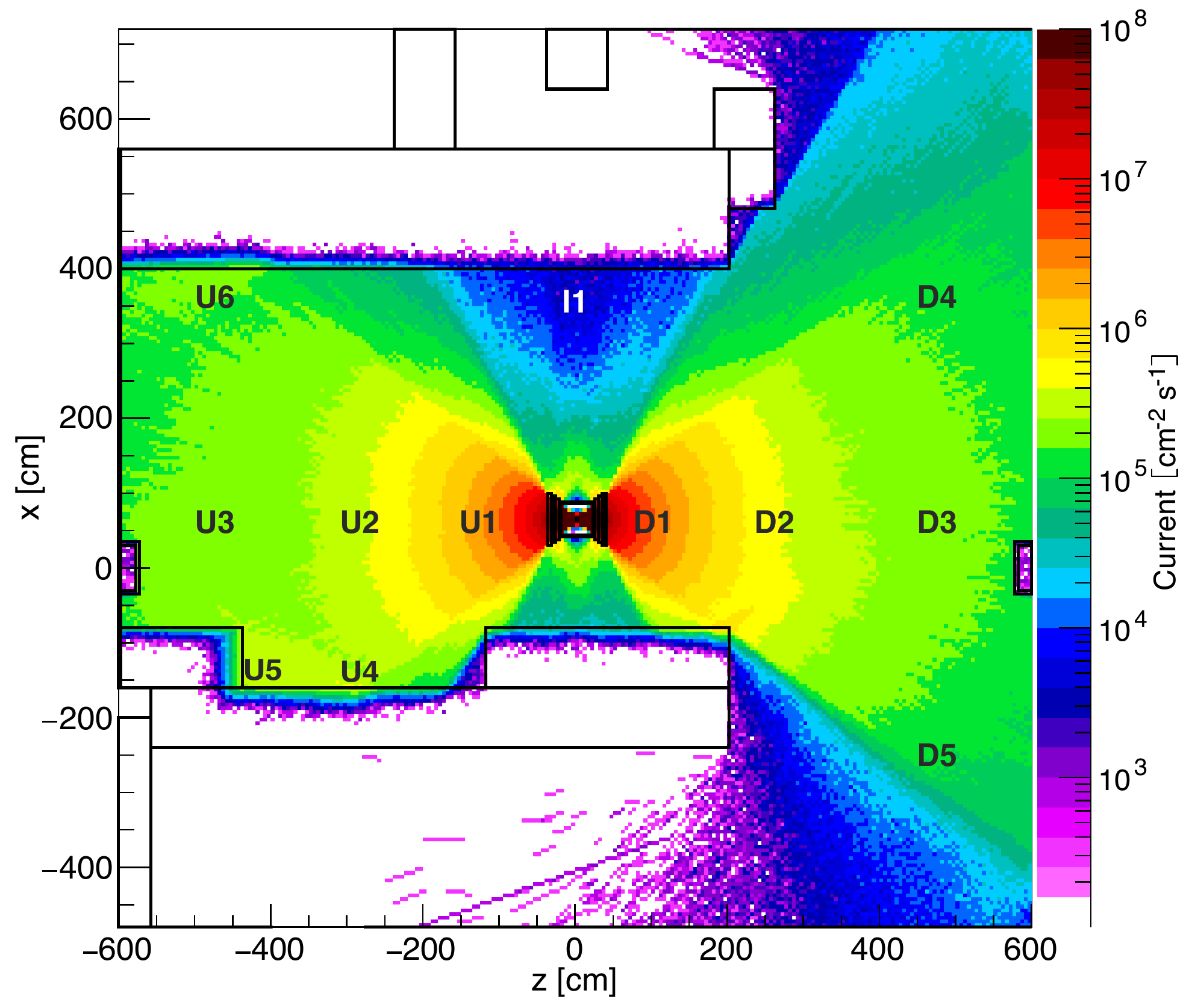}%
}
\subfloat[200~keV $<$ E $\le$ 300~keV \label{fig: Current_DS_US_x_200_300}]{
\includegraphics[width=.49\textwidth]{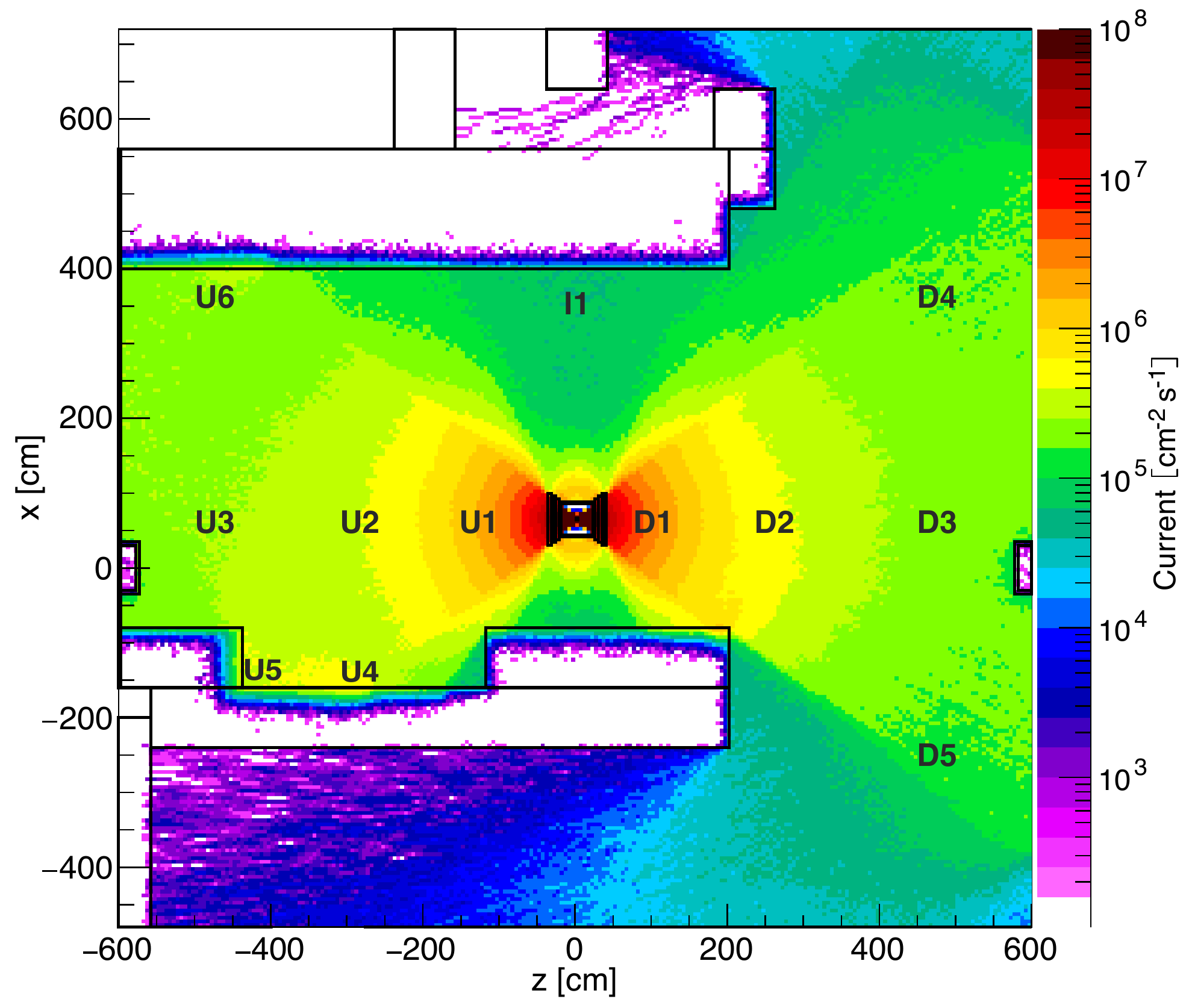}%
}
\\
\subfloat[100~keV $<$ E $\le$ 200~keV \label{fig: Current_DS_US_x_100_200}]{
\includegraphics[width=.49\textwidth]{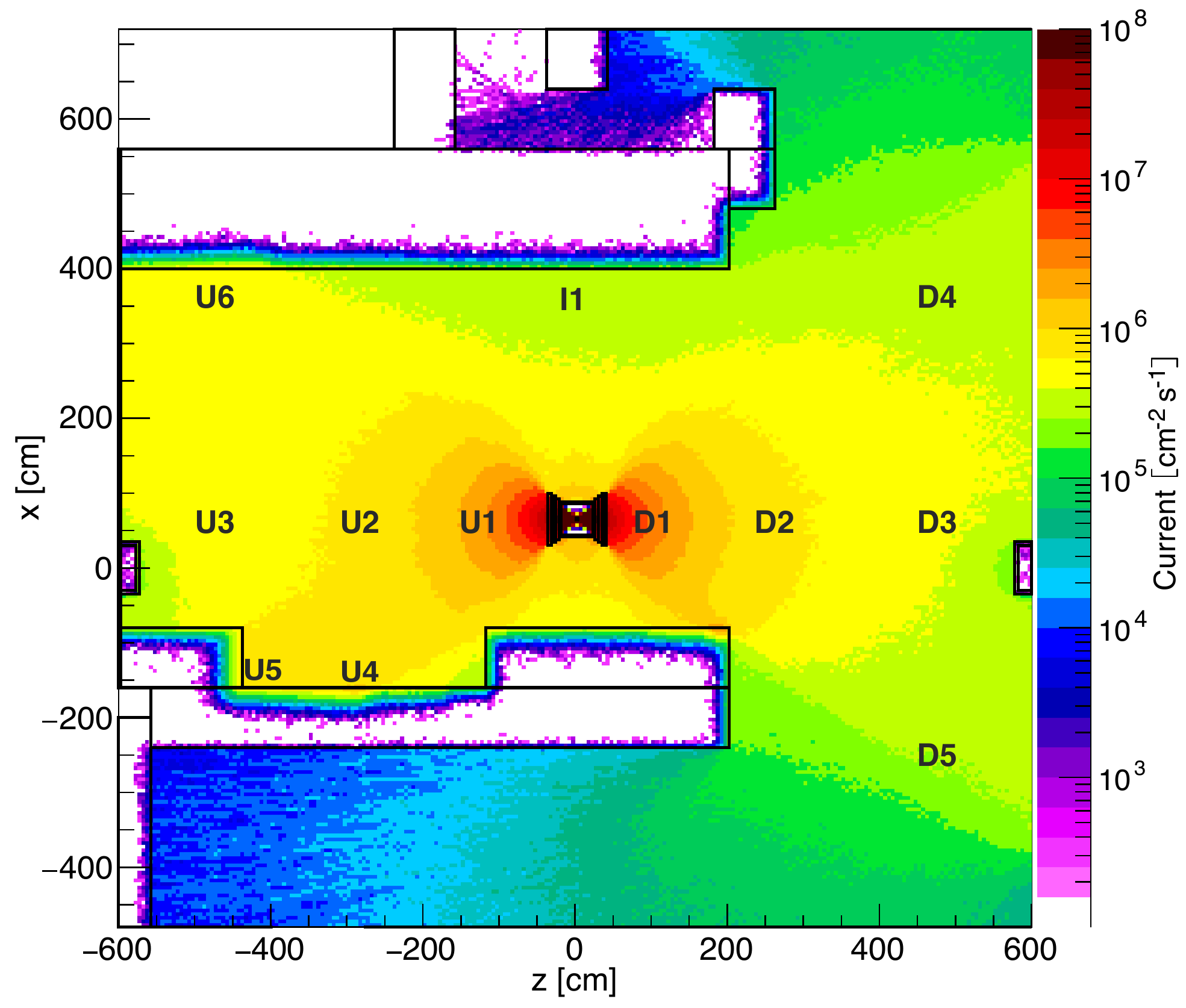}%
}
\subfloat[0~keV $<$ E $\le$ 100~keV \label{fig: Current_DS_US_x_000_100}]{
\includegraphics[width=.49\textwidth]{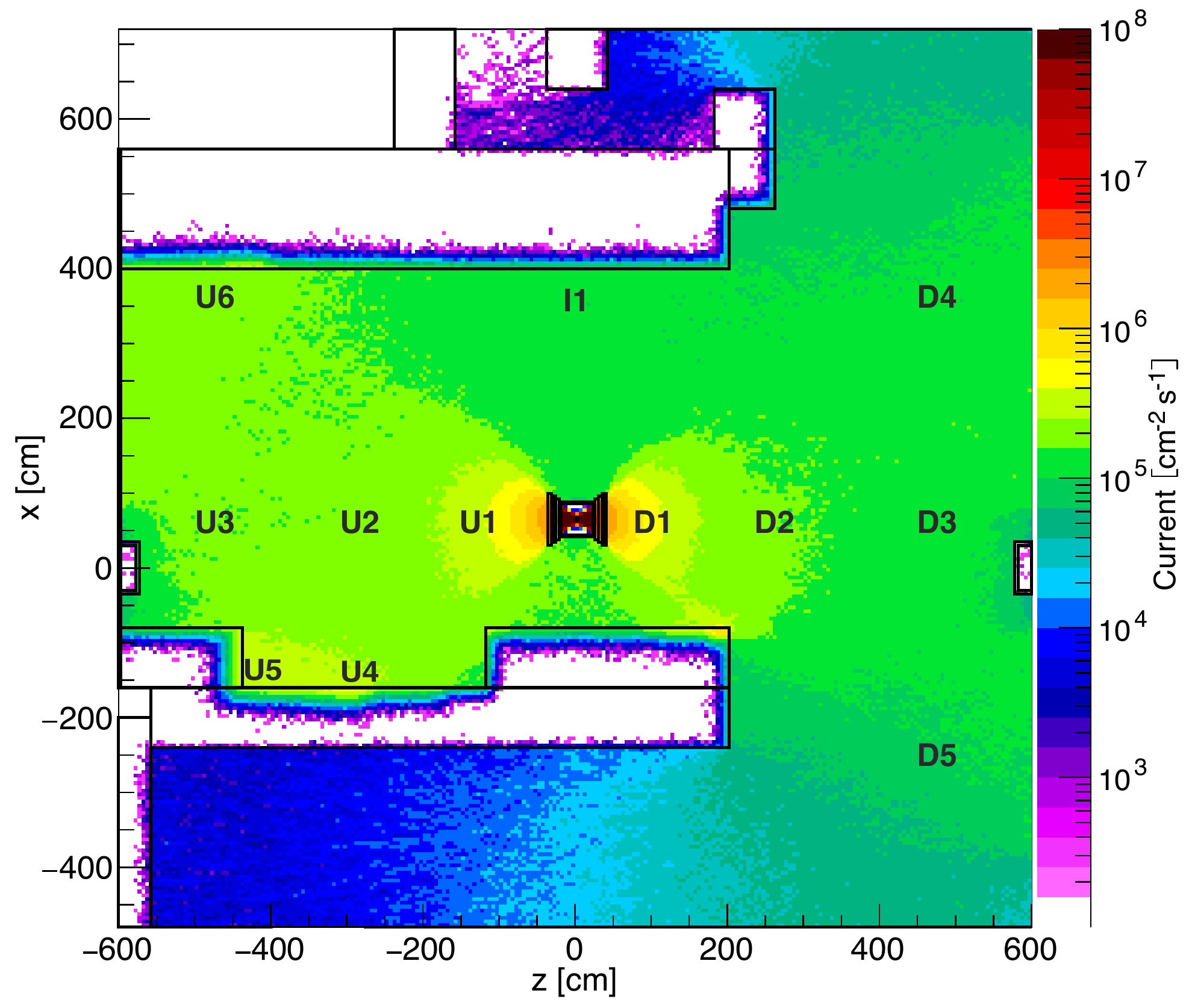}%
}
\caption{Simulated unattenuated current of photons in xz plane at y=0 m. The contributions of photons between a) 300~keV and 400~keV, b) 200~keV and 300~keV, c) 100~keV and 200~keV and d) 0~keV and 100~keV are shown. Photons with energy between 100 keV and 200 keV form the strongest contribution to the photon current below 400 keV.}
\label{fig: Simulated current 2}
\end{figure}

Figures~\ref{fig: Simulated current 1} and \ref{fig: Simulated current 2} depict the photon current at the height of the source in the xz plane of the facility. Thanks to the angular correction filters, the current of photons between 600~keV and 662~keV is uniform at given z value within the $\pm$~37$^\circ$ opening as displayed in Figure~\ref{fig: Current_DS_US_x_600_667}. This does not apply to the current of photons with energies below 500~keV. At a given z coordinate, the further away in x direction from the source a location is, the smaller the current. Due to the contribution from lower energy photons, the total current in Figure~\ref{fig: Current_DS_US_x} also shows some dependence from the x coordinate. In location I1, photons with an energy between 100~keV and 200~keV form the strongest contribution to the photon current as shown in Figure~\ref{fig: Current_DS_US_x_100_200}, whereas photons with an energy above 400~keV do not occur in this location. In the locations U4, U5 and U6, the current is somewhat enhanced (Figures~\ref{fig: Current_DS_US_x_200_300} to \ref{fig: Current_DS_US_x_000_100}) due to scattering of photons from the nearby walls, since the upstream zone of the GIF++ facility is narrower and has an 80~cm lower roof than the downstream zone. The absolute contribution of these scattered photons to the total current is nevertheless relatively small. Table~\ref{table: Simulated current} lists the contribution of the different energy ranges to the total photon current. Inside the $\pm$~37$^\circ$ wide irradiation area unattenuated 662~keV photons contribute between 33$\%$ and 54$\%$ of the total current. Directly in front of the irradiator about half of the photons are unattenuated photons. 

\begin{table}[htbp]
\centering
\scriptsize
\begin{threeparttable}
\begin{tabular}[htbp]{ l|r|r|r|r|r|r|r|r|r|r|}
\cline{2-11}
& \multicolumn{1}{ |c|}{\rot{90}{\shortstack[c]{0-100\\keV}}} & \multicolumn{1}{ |c|}{\rot{90}{\shortstack[c]{100-200\\keV}}} & \multicolumn{1}{ |c|}{\rot{90}{\shortstack[c]{200-300\\keV}}} & \multicolumn{1}{ |c|}{\rot{90}{\shortstack[c]{300-400\\keV}}} & \multicolumn{1}{ |c|}{\rot{90}{\shortstack[c]{400-500\\keV}}} & \multicolumn{1}{ |c|}{\rot{90}{\shortstack[c]{500-600\\keV}}}
& \multicolumn{1}{ |c|}{\rot{90}{\shortstack[c]{600-662\\keV}}} & \multicolumn{1}{ |c|}{\rot{90}{\shortstack[c]{0-662\\keV}}} & \multicolumn{2}{ |c|}{\rot{90}{\shortstack[c]{\phantom{0}661-662\phantom{0}\\keV}}}\\
\hline
\multicolumn{1}{ |l|}{\textbf{Position}} & \multicolumn{9}{ |c|}{[10$^6 cm^{-2} s^{-1}$]} & \multicolumn{1}{ |c|}{$\%$} \\
%\multicolumn{1}{ |l|}{\textbf{Pos.}} & $cm^{-2} s^{-1}$ & $cm^{-2} s^{-1}$ & $cm^{-2} s^{-1}$ & $cm^{-2} s^{-1}$ & $cm^{-2} s^{-1}$ & $cm^{-2} s^{-1}$
%& $cm^{-2} s^{-1}$ & $cm^{-2} s^{-1}$& $\%$& $cm^{-2} s^{-1}$\\
\hline
\multicolumn{1}{ |l|}{D1}       & 0.36 & 2.92 & 2.44 & 3.29  & 5.18  & 5.88   & 23.8 & 43.9 & 20.0 & 45.5  \\
\multicolumn{1}{ |l|}{D2}       & 0.17 & 0.72 & 0.45 & 0.35  & 0.67  & 0.83   & 3.34 & 6.52 & 2.79 & 42.8  \\
\multicolumn{1}{ |l|}{D3}       & 0.14 & 0.47 & 0.22 & 0.20  & 0.23  & 0.29   & 1.14 & 2.70 & 0.95 & 35.2  \\
\multicolumn{1}{ |l|}{D4}       & 0.12 & 0.38 & 0.19 & 0.13  & 0.14  & 0.15   & 1.28 & 2.39 & 1.17 & 48.7  \\
\multicolumn{1}{ |l|}{D5}       & 0.11 & 0.32 & 0.17 & 0.13  & 0.15  & 0.15   & 1.29 & 2.32 & 1.17 & 50.3  \\
\multicolumn{1}{ |l|}{I1}       & 0.13 & 0.35 & 0.08 & 0.008 & 0.005 & 0.0002 & 0.00 & 0.57 & 0.00 & 0.00  \\
\multicolumn{1}{ |l|}{U1}       & 0.39 & 2.96 & 2.44 & 3.30  & 5.14  & 5.89   & 23.8 & 43.9 & 20.0 & 45.5  \\
\multicolumn{1}{ |l|}{U2}       & 0.21 & 0.81 & 0.41 & 0.34  & 0.66  & 0.83   & 3.38 & 6.64 & 2.82 & 42.6  \\
\multicolumn{1}{ |l|}{U3}       & 0.19 & 0.57 & 0.28 & 0.20  & 0.23  & 0.28   & 1.15 & 2.91 & 0.96 & 33.1  \\
\multicolumn{1}{ |l|}{U3$_{a}$} & 0.18 & 0.54 & 0.27 & 0.21  & 0.27  & 0.29   & 1.15 & 2.90 & 0.96 & 33.3  \\
\multicolumn{1}{ |l|}{U3$_{b}$} & 0.14 & 0.47 & 0.26 & 0.19  & 0.26  & 0.31   & 1.18 & 2.82 & 1.01 & 36.0  \\
\multicolumn{1}{ |l|}{U4}       & 0.29 & 0.95 & 0.49 & 0.27  & 0.33  & 0.35   & 3.82 & 6.51 & 3.54 & 54.4  \\
\multicolumn{1}{ |l|}{U5}       & 0.34 & 0.86 & 0.45 & 0.32  & 0.29  & 0.26   & 1.71 & 4.23 & 1.52 & 35.9  \\
\multicolumn{1}{ |l|}{U6}       & 0.22 & 0.56 & 0.29 & 0.19  & 0.15  & 0.15   & 1.28 & 2.84 & 1.16 & 40.9  \\ 

\hline
\end{tabular}
\caption{Overview of simulated photon current [10$^6 cm^{-2} s^{-1}$]: downstream and upstream fully open. These are the highest currents available.}
\label{table: Simulated current}
\end{threeparttable}
\end{table}

\section{Conclusions}
\label{sec: Conclusions}
The new GIF++ radiation facility, operational since spring 2015, features an intense source of 662~keV photons. The irradiator has two openings of $\pm$~37$^\circ$ oriented towards a downstream and an upstream irradiation area. A high-energy muon beam passes close to the irradiator and permits to study detector performance while being irradiated at adjustable high rates. Each irradiation zone is equipped with a versatile filter system permitting to attenuate the current of 662~keV photons in 24 steps by up to a factor 4.6$\times$10$^{4}$. To improve uniformity over large planar detectors, integrated angular correction filters serve to ensure that the current of the unattenuated photons depends predominantly on the z coordinate. Lower energy photons, from interactions in irradiator and in surrounding material, contribute in a complex way to the radiation field. The photon current in the whole bunker has been simulated with Geant4. Photon currents of up to 5$\times$10$^{7}$ photons/(cm$^2$ s)  are available at a distance of 1~m from the source. Between 33$\%$ and 54$\%$ of the photon current comes from unattenuated photons with an energy of 662~keV, the remainder are photons that lost energy mostly through Compton scattering in the lead of the filters, in the steel of the facility floor or in the concrete of walls and roof. RADMON measurements of the absorbed dose in air have been compared with simulations. The measurements agree to 12$\%$ with the simulations, and are well within expected uncertainties. The same agreement is also found for the Automess 6150AD-15 measurements in locations close to the source. The maximum absorbed dose rate in air amounts to 2.2 Gy/h directly in front of the irradiator, at a distance of 50 cm from the center of the source.

\section{Acknowledgments}
\label{sec:Acknowledgements}
DP would like to acknowledge the support of the EU Horizon 2020 framework, BrightnESS project 676548.

\bibliographystyle{elsarticle-num}

\end{document}